\documentclass[preprint,showpacs,onecolumn,nofootinbib]{revtex4}
\pdfoutput=1
\usepackage[colorlinks=true,linkcolor=blue,urlcolor=blue,filecolor=black,citecolor=red,pdfstartview=FitV,pdftitle={},pdfsubject={},pdfkeywords={},pdfpagemode=None,bookmarksopen=true]{hyperref}
\usepackage{graphicx}
\usepackage{amsfonts}
\usepackage{amssymb,ulem}
\usepackage{color}%
\usepackage{dcolumn}
\usepackage{amsmath}
\usepackage{epsfig}
\usepackage{graphics}
\usepackage[active]{srcltx}
\usepackage{epstopdf}
\usepackage{shuffle}
\usepackage[utf8]{inputenc}
\usepackage{bbm}

\newcommand{\be}{\begin{equation}}
\newcommand{\ee}{\end{equation}}
\newcommand{\bea}{\begin{eqnarray}}
\newcommand{\eea}{\end{eqnarray}}
\newcommand{\bean}{\begin{eqnarray*}}
\newcommand{\eean}{\end{eqnarray*}}
\newcommand{\nn}{\nonumber \\}

\def\W #1{\widetilde{#1}}

\def\eref#1{(\ref{#1})}

\def\Label#1{\label{#1}%
  \smash{\hbox to0pt{\raise1ex\hbox{\tiny[#1]}\hss}}}

\begin{document}

\baselineskip=0.6 cm
\title{Note on scalar-graviton and scalar-photon-graviton amplitudes}
\author{Kang Zhou}
\email{zhoukang@yzu.edu.cn}
\author{Guo-Jun Zhou}
\email{zhou1750819726@163.com}

\affiliation{Center for Gravitation and Cosmology, College of Physical Science and Technology, Yangzhou University, Yangzhou, 225009, China}

\begin{abstract}
\baselineskip=0.6 cm
In this short note, we propose an algorithm based on the expansions of amplitudes, the dimensional reduction technic and the differential operators, to calculate the tree level scalar-graviton amplitudes with two massive scalars, as well as the tree level scalar-photon-graviton amplitudes with two massive scalars and one photon. While applying the unitarity method, these amplitudes are necessary inputs for the calculation of post-Newtonian and post-Minkowskian expansions in general relativity for two massive charged objects interact with gravity and electromagnetic field.

\end{abstract}


\maketitle


\section{Introduction}
\label{secintro}

The progress in the study of scattering amplitudes in the past decade has revealed
deep physical insights into the quantum field theory, and inspired efficient methods for practical calculation.
Recently, an effort has emerged to
connect the amplitudes program to the physics of gravitational waves, which were discovered at LIGO/Virgo \cite{Abbott:2016blz,TheLIGOScientific:2017qsa}.
More explicitly, the modern tools for calculating scattering amplitudes at loop level provide a powerful new way to evaluate
post-Newtonian and post-Minkowskian expansions in classical general relativity, give rise to effective two-body Hamiltonians
\cite{Cachazo:2017jef,Guevara:2017csg,Damour:2017zjx,Bjerrum-Bohr:2018xdl,Levi:2018nxp,Cheung:2018wkq,Chung:2018kqs,Bern:2019nnu,Bern:2019crd,
Antonelli:2019ytb,Cristofoli:2019neg,KoemansCollado:2019ggb,Maybee:2019jus,Bjerrum-Bohr:2019nws}.

In this interesting new direction, an essential step is to calculate the scattering of two massive objects interact with gravitons, at $n$-loop order. To achieve the goal, the tree level amplitudes with two massive scalars and $n+1$ gravitons are required,
due to the so called unitarity method. Among several methods proposed to calculate these tree amplitudes, one remarkable approach based on the CHY formalism was suggested by Naculich \cite{Naculich:2015zha}. The advantage of this method is that the CHY formalism is valid in arbitrary space-time dimensions \cite{Cachazo:2013gna,Cachazo:2013hca,Cachazo:2013iea,Cachazo:2014nsa,Cachazo:2014xea}, thus the obtained results are more suitable for the dimensional regularization at intermediate steps. More recently, another more efficient way, makes use of the CHY formalism as well as the double-cover construction \cite{Gomez:2016bmv,Cardona:2016bpi,Bjerrum-Bohr:2018lpz,Gomez:2018cqg,Bjerrum-Bohr:2018jqe,Gomez:2019cik}, was proposed by Bjerrum-Bohr, Cristofoli, Damgaard and Gomez \cite{Bjerrum-Bohr:2019nws}. Using their method, one can evaluate the desired amplitudes recursively, thus avoid the treatments of contour integrals in the CHY formalism.

\begin{figure}
  \centering
  \includegraphics[width=10cm]{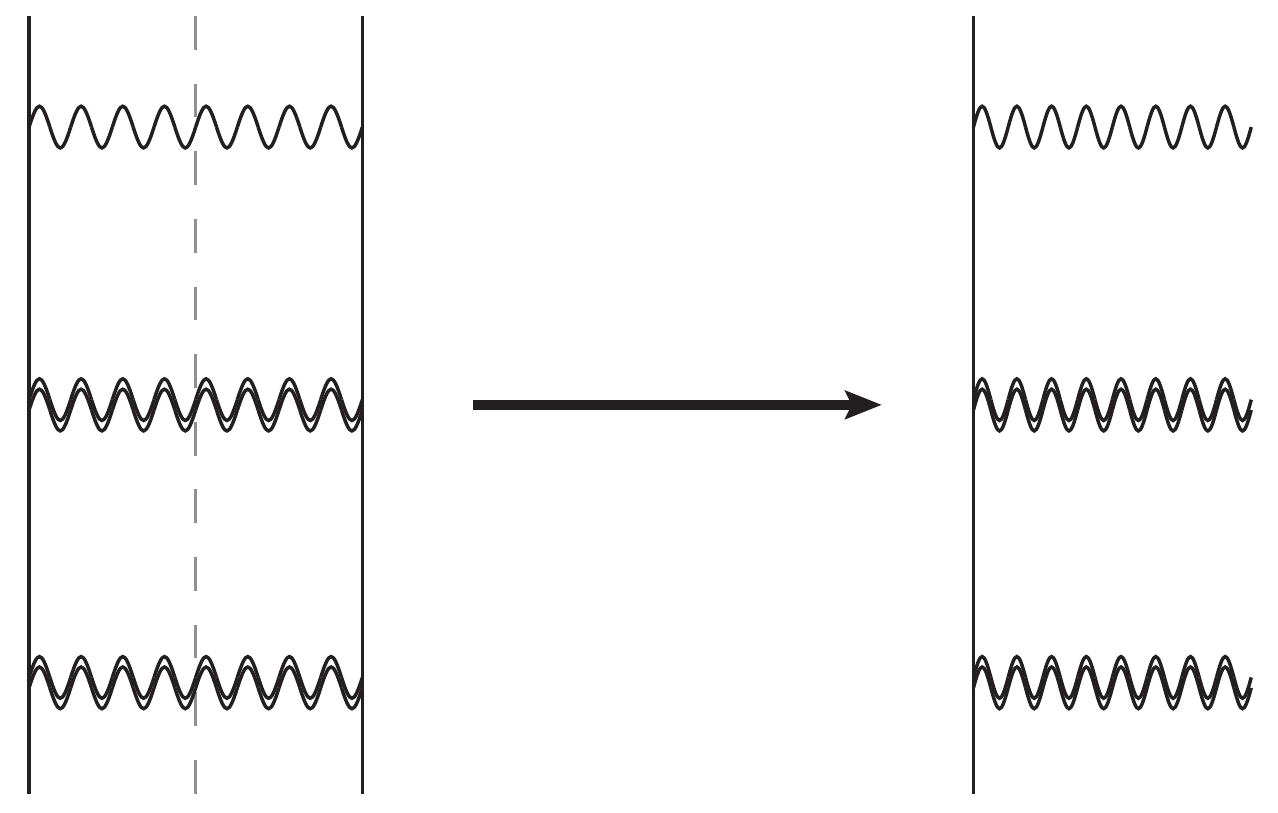}
  \caption{Unitarity method for the scattering of two massive particles interact with gravitons and photons, at loop level. The straight lines represent the charged massive objects, the wavy lines stand for photons, and the double-wavy lines stand for gravitons.}\label{cut}
\end{figure}

One can also consider the classical system including two charged massive objects, such as two charged black holes, interact to each other through both gravity and electromagnetic field. Via the idea similar as that described above, one can seek the effective two-body Hamiltonian by calculating the scattering of two massive particles interact with gravitons and photons, at loop level. Then the unitarity method motivated us to consider the tree level amplitudes whose external states including two massive scalars, gravitons, and photons, as shown in Fig.\ref{cut}.

In this short note, we propose an algorithm based on the expansions of amplitudes \cite{Stieberger:2016lng,Schlotterer:2016cxa,Chiodaroli:2017ngp,DelDuca:1999rs,Nandan:2016pya,
delaCruz:2016gnm,Fu:2017uzt,Teng:2017tbo,Du:2017kpo,Du:2017gnh,Feng:2019cbe,Hu:2019qdq,Zhou:2019mbe}, the differential operators constructed by Cheung, Shen and Wen \cite{Cheung:2017ems,Zhou:2018wvn,Bollmann:2018edb}, or equivalently the dimensional reduction technic \cite{Naculich:2015zha,Cachazo:2014xea}, to calculate the tree level scalar-graviton (SG) amplitudes with two massive scalars, as well as the tree level scalar-photon-graviton (SPG) amplitudes with two massive scalars and one photon.
The first step of the algorithm is to compute the gravity (GR) amplitude with two massive gravitons by expanding it to the bi-adjoint scalar (BAS) amplitudes. The methods for calculating basis and coefficients in the expansion will be discussed in \S\ref{secalgo}.
The SG amplitude with two massive scalars can be generated from the GR amplitude by converting two massive gravitons to scalars, via the dimensional  reduction manipulation, or applying the differential operators. The SPG amplitude with two massive scalars and one photon, can be obtained by transmuting the GR amplitude to the single trace Einstein-Yang-Mills (EYM) amplitude with two massive and one massless gluons, then converting two massive gluons to scalars, and identifying the remaining massless gluon as a photon.
Our approach is also regardless of the space-time dimensions, and can be easily implemented in MATHEMATIC. As will be explained, the SPG amplitudes with more than one photons is hard to be computed by the method developed in this paper. We leave this problem to the future work.

The remainder of this note is organized as follows. In \S\ref{secalgo}, we introduce our algorithm in detail. In \S\ref{secSG}, we compute the SG examples via this algorithm.
In \S\ref{secSPG}, the SPG examples are considered. The full expressions of the $5$-point SG and SPG amplitudes are exhibited in Appendix \S\ref{secappen1}.

\section{The algorithm}
\label{secalgo}

Our algorithm is to evaluate tree level GR amplitudes with two massive gravitons via the expansion obtained in \cite{Zhou:2019mbe}, then convert gravitons to massive scalars and massless photons through the dimensional reduction procedure \cite{Naculich:2015zha,Cachazo:2014xea}, or applying differential operators \cite{Cheung:2017ems,Zhou:2018wvn,Bollmann:2018edb}, to obtain the desired scalar-graviton (SG) or scalar-photon-graviton (SPG) amplitudes. In subsection \S\ref{subsec-calcu-GR}, we will discuss the calculation of GR amplitudes, including the technics for computing coefficients in the expansion, as well as the BAS amplitudes which serve as basis. In subsection \S\ref{subsec-conver}, we will study how to transmute gravitons to scalars or photons.

\subsection{Calculating GR amplitude}
\label{subsec-calcu-GR}

When all external legs are massless, the tree level GR amplitude in $D$ space-time dimensions can be expanded to BAS amplitudes in $D$ dimensions, in the following double copy formula \cite{Zhou:2019mbe}
\bea
{\cal A}_{\rm GR}(1,\cdots,n)=\sum_{\sigma}\sum_{\sigma'}\,C^\epsilon(\sigma){\cal A}_{\rm BAS}(1\sigma n|1\sigma'n)C^{\W\epsilon}(\sigma')\,,~~~~\label{exp-GR}
\eea
where $\sigma,\sigma'\in S_{n-2}$ are permutations of $n-2$ elements in $\{2,\cdots,n-1\}$. Each basis ${\cal A}_{\rm BAS}(1\sigma n|1\sigma'n)$ carries two color-orderings $(1\sigma n)$ and $(1\sigma'n)$. We have chosen $(1\cdots n)$ to denote legs with a color-ordering, and $(1,\cdots,n)$ to denote legs without any color-ordering. This convention will be used to all amplitudes, throughout this paper. Coefficients $C^\epsilon(\sigma)$ and $C^{\W\epsilon}(\sigma')$ are just Bern-Carrasco-Johansson (BCJ) numerators for Yang-Mills amplitudes \cite{Bern:2008qj,Chiodaroli:2014xia,Johansson:2015oia,Johansson:2019dnu}. The superscripts $\epsilon$ and $\W\epsilon$ indicate the dependence on polarization vectors in two sets $\{\epsilon_i\}$ and $\{\W\epsilon_i\}$, respectively.
The GR amplitude carries two independent sets of polarization vectors is understood in a generalized version of the gravity theory, i.e., Einstein gravity couples to a dilaton and $2$-form.
It can be reduced to the amplitude for pure Einstein gravity, by setting all $\W\epsilon_i=\epsilon_i$. For our purpose, we want to seek SP and SPG amplitudes including gravitons for pure Einstein gravity. However, at intermediate steps, we will keep $\{\epsilon_i\}$ and $\{\W\epsilon_i\}$ to be different, to manifest the double copy structure. This structure ensures that the dimensional reduction procedure, or differential operators, act only on one piece, while keeping another one un-altered. After finishing these manipulations, we will turn all $\W\epsilon_i$ to $\epsilon_i$, to exclude the contributions from the dilaton and $2$-form.

Now we use the CHY formula to explain that such expansion is also correct when two external legs are massive (both GR and BAS amplitudes contain two massive legs).
In the CHY formula, the GR and BAS amplitudes for massless external legs arise as the contour integrals \cite{Cachazo:2013gna,Cachazo:2013hca,Cachazo:2013iea,Cachazo:2014nsa,Cachazo:2014xea}
\bea
{\cal A}^{\epsilon,\W\epsilon}_{\rm GR}(1,\cdots,n)&=&\int\,d\mu_n\,{\bf Pf}'\Psi_n^\epsilon{\bf Pf}'\Psi_n^{\W\epsilon}\,,\nn
{\cal A}_{\rm BAS}(1\sigma n|1\sigma'n)&=&\int\,d\mu_n\,PT_n(1\sigma n)
PT_n(1\sigma'n)\,,
\eea
with the universal measure $d\mu_n$. From the CHY point of view, expanding the GR amplitude to BAS ones can be understood as the expansions
\bea
{\bf Pf}'\Psi_n^\epsilon=\sum_\sigma\,C^\epsilon(\sigma)PT_n(1\sigma n)\,,~~~~{\bf Pf}'\Psi_n^{\W\epsilon}=\sum_{\sigma'}\,C^{\W\epsilon}(\sigma')PT_n(1\sigma'n)\,.~~~~\label{exp-pfa}
\eea
When two external legs are massive, the Parke-Taylor factors $PT_n(1\sigma n)$ and $PT_n(1\sigma'n)$ will not be altered \cite{Naculich:2015zha,Lam:2019mfk,Bjerrum-Bohr:2019nws}.
On the other hand, if one choose the removed rows and columns for the reduced Pfaffians ${\bf Pf}'\Psi_n^{\epsilon}$ and ${\bf Pf}'\Psi_n^{\W\epsilon}$ to be two massive legs, two reduced Pfaffians are also un-modified \cite{Naculich:2015zha,Lam:2019mfk,Bjerrum-Bohr:2019nws}. Thus one can conclude that the expansions in
\eref{exp-pfa} still holds.
When two external legs are massive, the measure part $d\mu_n$ will be altered due to the modification for the so called scattering equations \cite{Naculich:2015zha,Lam:2019mfk,Bjerrum-Bohr:2019nws}. But it is still universal if the massive legs have the same nodes, i.e., if the $i^{\rm th}$ and $j^{\rm th}$ legs are massive gravitons for the GR amplitude, $i$ and $j$ also denote massive scalars for the BAS amplitudes. Combining this fact with the expansions in \eref{exp-pfa}, we arrive at the conclusion that the expansion in \eref{exp-GR} is valid for amplitudes with two massive legs, if the massive legs have the same nodes.

Thus, one can use the expansion in \eref{exp-GR} to compute the GR amplitude. To do so, one need to calculate the BAS amplitudes which serve as the basis, as well as the coefficients $C^{\epsilon}(\sigma)$ and $C^{\W\epsilon}(\sigma')$ in the expansion.

\begin{figure}
  \centering
  \includegraphics[width=6cm]{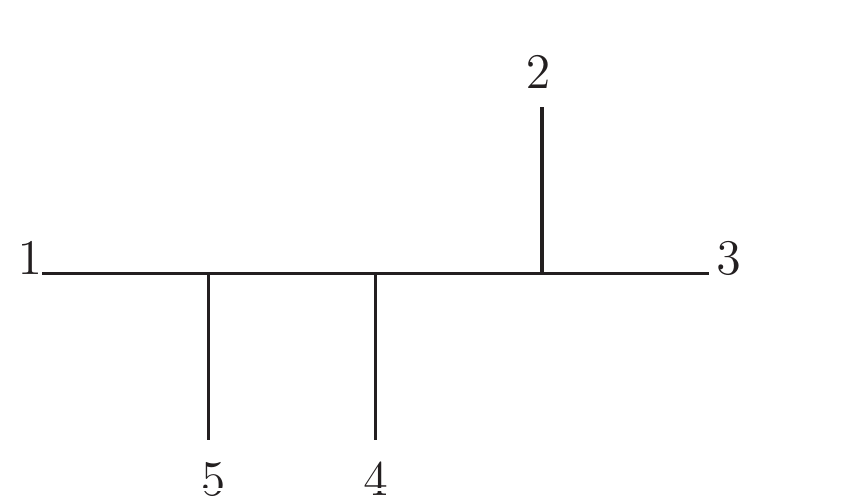}
   \includegraphics[width=6cm]{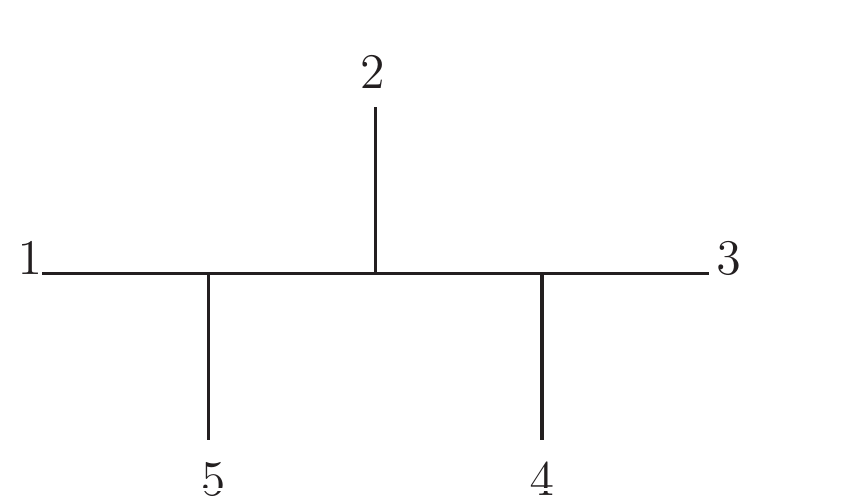}  \\
  \caption{Two $5$-point diagrams}\label{5p}
\end{figure}

To calculate BAS amplitudes with two massive legs, we employ the method proposed by Cachazo, He and Yuan in \cite{Cachazo:2013iea}.
For a BAS amplitude whose double color-orderings are given, this method provides the corresponding Feynman diagrams as well as the overall sign directly from the color-orderings.
We find this method to be effective, since none of individual Feynman diagrams for pure scalar amplitudes have any gauge redundancy, and the double color-orderings
reduce the number of diagrams greatly. To illustrate, let us consider the $5$-point example ${\cal A}_{\rm BAS}(12345|14235)$.
In Fig.\ref{5p}, the first diagram satisfies both two color-orderings $(12345)$ and $(14235)$, while the second one satisfies the ordering
$(12345)$ but not $(14235)$. Thus, the first diagram is allowed by the double color-orderings $(12345|14235)$, while the second one is not. It is easy to see other diagrams are also forbidden by the ordering $(14235)$, thus the first diagram in Fig.\ref{5p} is the only diagram contributes to the amplitude
${\cal A}_{\rm BAS}(12345|14235)$. Thus, in this example, the number of Feynman diagrams is markedly reduced by two color-orderings.

\begin{figure}
  \centering
   \includegraphics[width=5cm]{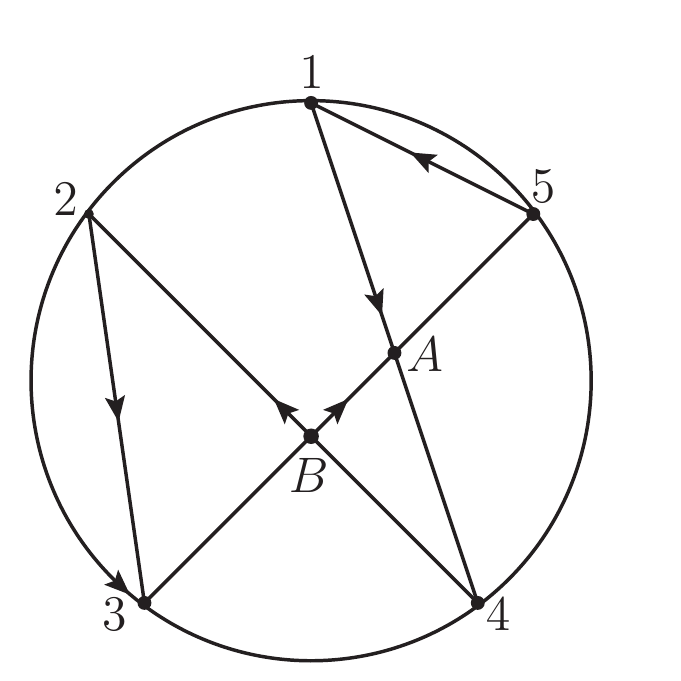} \\
  \caption{Diagram for ${\cal A}_{\rm BAS}(1,2,3,4,5|1,4,2,3,5)$}\label{dis14235}
\end{figure}

\begin{figure}
  \centering
  \includegraphics[width=10cm]{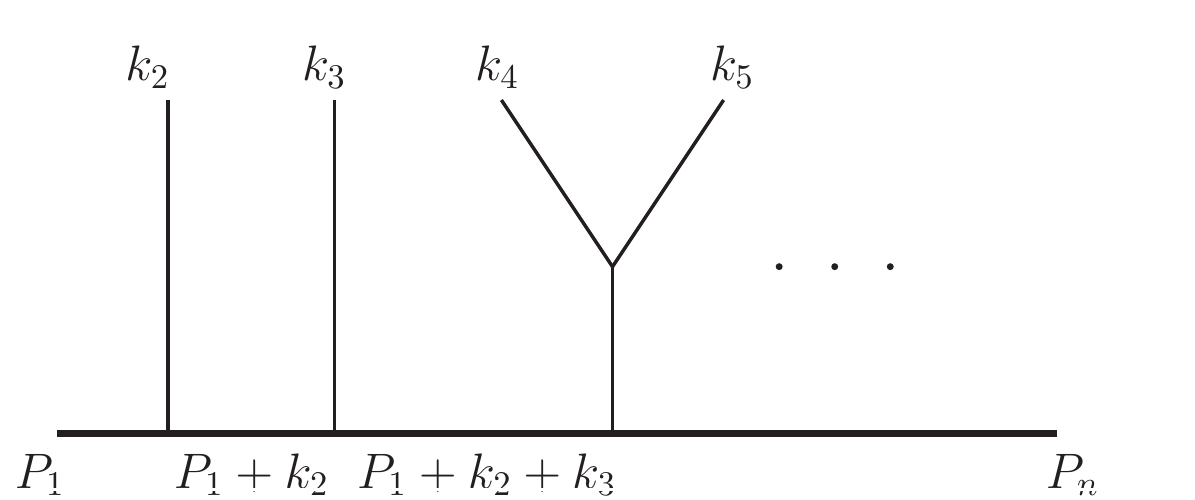}
  \caption{Typical diagram for BAS amplitudes with two massive scalars, the bold line represents the massive particle}\label{massive-pro}
\end{figure}

The Feynman diagrams for a given BAS amplitude can be obtained via a systematic diagrammatic rule. For the above example, one can draw a disk diagram as follows.
\begin{itemize}
\item Draw points on the boundary of the disk according to the first ordering $(12345)$.
\item Draw a loop of line segments which connecting the points according to the second ordering $(14235)$.
\end{itemize}
The obtained disk diagram is shown in Fig.\ref{dis14235}. From the diagram, one can see that two orderings share the boundaries $\{1,5\}$ and $\{2,3\}$\footnote{Here we use $\{\cdots\}$ rather than $(\cdots)$ to emphasize that two co-boundaries are regardless of orderings.}. These co-boundaries
indicate channels ${1\over s_{15}}$ and ${1\over s_{23}}$, therefore the first Feynman diagram in Fig.\ref{5p}. Then the BAS amplitude ${\cal A}_{\rm BAS}(12345|14235)$ can be computed as
\bea
{\cal A}_{\rm BAS}(12345|14235)={1\over s_{23}}{1\over s_{15}}\,,
\eea
up to an overall sign. In this paper the kinematic variables $s_{ij\cdots k}$ are defined as
\bea
s_{ij\cdots k}=(k_i+k_j+\cdots+k_k)^2-(k_i^2+k_j^2+\cdots+k_k^2)\,.
\eea
The advantage of this definition is that the propagators expressed by ${1\over s_{ij\cdots k}}$ are valid for both massless
amplitudes and amplitudes with two massive external legs which we are interested, as can be seen through the following discussion. The typical Feynman diagram for BAS amplitudes with
two massive scalars is shown in Fig.\ref{massive-pro}. Legs $1$ and $n$ are assumed to be massive, with $P_1^2=P_n^2=m^2$. To distinguish them from massless particles,
we have introduced $P_1$ and $P_n$ to denote two massive momenta. From Fig.\ref{massive-pro}, one can observe that each massive virtual particle
provides the propagator in the form
\bea
{1\over(P_1+k_2+k_3+\cdots)^2-m^2}={1\over(P_1+k_2+k_3+\cdots)^2-P_1^2}={1\over s_{123\cdots}}\,.
\eea
Thus, the formula ${1\over s_{ij\cdots k}}$ is the correct expression for any propagator in the case of two massive scalars.
Notice that when two massive legs belong to ${\pmb \alpha}$ where ${\pmb \alpha}$ is a subset of external legs $\{1,\cdots,n\}$,
we must use $s_{\{1,\cdots,n\}\setminus{\pmb\alpha}}$ rather than $s_{\pmb\alpha}$, since the second one can not reproduce
the correct propagator corresponding to the channel ${\pmb \alpha}$.
For other cases, it is easy to prove that $s_{\pmb\alpha}=s_{\{1,\cdots,n\}\setminus{\pmb\alpha}}$.

\begin{figure}
  \centering
   \includegraphics[width=5cm]{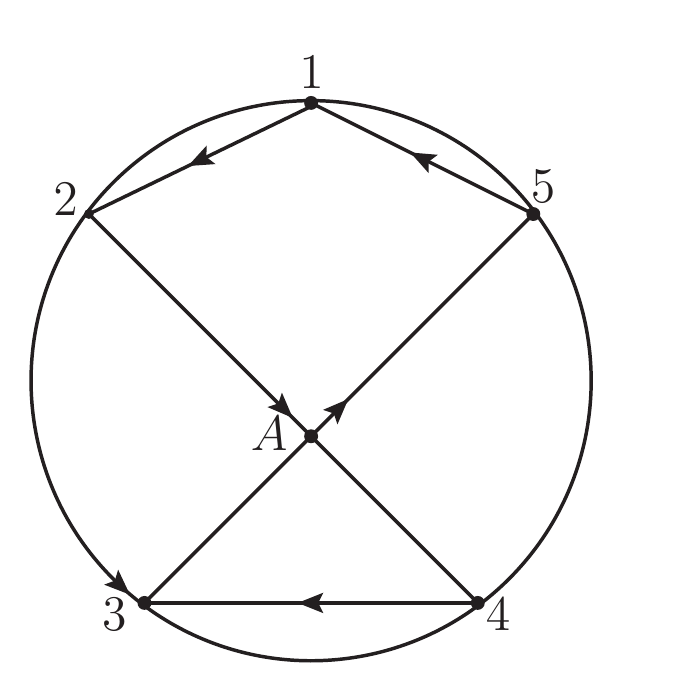} \\
  \caption{Diagram for ${\cal A}_{\rm BAS}(1,2,3,4,5|1,2,4,3,5)$}\label{dis12435}
\end{figure}

As another example, let us consider the BAS amplitude ${\cal A}_{\rm BAS}(12345|12435)$. The corresponding disk diagram is shown in
Fig.\ref{dis12435}, and one can see two orderings have co-boundaries $\{3,4\}$ and $\{5,1,2\}$. The co-boundary $\{3,4\}$ indicates the channel ${1\over s_{34}}$. The co-boundary $\{5,1,2\}$ indicates the channel ${1\over s_{512}}$, as well as sub-channels ${1\over s_{12}}$ and ${1\over s_{51}}$
(${1\over s_{52}}$ is forbidden by both two orderings). Notice that the channel ${1\over s_{512}}$ is equivalent to ${1\over s_{34}}$. Using the above decomposition, one can calculate ${\cal A}_{\rm BAS}(12345|12435)$ as
\bea
{\cal A}_{\rm BAS}(12345|12435)={1\over s_{34}}\Big({1\over s_{12}}+{1\over s_{234}}\Big)\,,
\eea
up to an overall sign.

The overall sign can be fixed by the following rule.
\begin{itemize}
\item Each polygon with odd number of vertices contributes
a plus sign if its orientation is the same as that of the disk and a minus sign if opposite.
\item Each polygon with even number of vertices always contributes a minus sign.
\item Each intersection point contributes a minus sign.
\end{itemize}
We can apply this rule to the previous examples. In Fig.\ref{dis14235}, the polygons are three triangles, namely $51A$, $A4B$ and $B23$, which contribute $+$, $-$, $+$ respectively, while two intersection points $A$ and $B$ contribute two $-$. In Fig.\ref{dis12435}, the polygons are $512A$ and $A43$, which contribute two $-$, while the intersection point $A$ contributes $-$. Then we arrive at the full results
\bea
{\cal A}_{\rm BAS}(12345|14235)&=&-{1\over s_{23}}{1\over s_{234}}\,,\nn
{\cal A}_{\rm BAS}(12345|12435)&=&-{1\over s_{34}}\Big({1\over s_{12}}+{1\over s_{234}}\Big)\,.
\eea

The BCJ numerators $C^\epsilon(\sigma)$ and $C^{\W\epsilon}(\sigma')$ in \eref{exp-GR} can be obtained by the rule provided in \cite{Fu:2017uzt,Teng:2017tbo,Zhou:2019mbe}.
To explain this rule,
we chose a reference ordering $n\prec j_2\prec\cdots\prec j_n$,
with $n$ is fixed at the lowest position. We denote the reference ordering as $\pmb{\cal R}$, and denote the color-ordering $(1\sigma n)$
as $1\dot{<} \sigma_2\dot{<}\cdots\dot{<} \sigma_{n-1}\dot{<}n$.
Then, the so called ordered splittings for the ordering $(1\sigma n)$ can be constructed via the following procedure.
\begin{itemize}
\item At the first step, we construct all possible ordered subsets $\pmb{{\alpha}}_0=\{1,\alpha^0_2,\cdots,\alpha^0_{|0|-1},n\}$, which satisfy two conditions, (1) $\pmb{{\alpha}}_0\subset\{1,\cdots,n\}$, (2) $\alpha^0_2\dot{<}\alpha^0_3\dot{<}\cdots\dot{<}\alpha^0_{|0|-1}$, respecting to the color-ordering $1\dot{<} \sigma_2\dot{<}\cdots\dot{<} \sigma_{n-1}\dot{<}n$.
    Here $|i|$ stands for the length of the set $\pmb{\alpha}_i$.
    We call each ordered subset $\pmb{{\alpha}}_0$ a root.
\item For each root $\pmb{{\alpha}}_0$, we eliminate its elements in $\{1,\cdots,n\}$ and $\pmb{\cal R}$, resulting in a reduced set $\{1,\cdots,n\}\setminus\pmb{{\alpha}}_0$, and a reduce reference ordering $\pmb{\cal R}\setminus\pmb{{\alpha}}_0$. Suppose $R_1$ is the lowest element in the reduce reference ordering $\pmb{\cal R}\setminus\pmb{{\alpha}}_0$, we construct all possible ordered subsets $\pmb{{\alpha}}_1$ as $\pmb{{\alpha}}_1=\{\alpha_1^1,\alpha_2^1,\cdots,\alpha_{|1|-1}^1,R_1\}$, with $\alpha_1^1\dot{<}\alpha_2^1\dot{<}\cdots\dot{<}\alpha_{|1|-1}^1\dot{<}R_1$, regarding to the color-ordering.
\item By iterating the second step, one can construct $\pmb{{\alpha}}_2,\pmb{{\alpha}}_3,\cdots$, until $\pmb{\alpha}_0\cup\pmb{\alpha}_1\cup\cdots\cup\pmb{\alpha}_r=\{1,\cdots,n\}$.
\end{itemize}
Each ordered splitting is given as an ordered set $\{\pmb{{\alpha}}_0,\pmb{{\alpha}}_1,\cdots,\pmb{{\alpha}}_r\}$, where ordered sets
$\pmb{{\alpha}}_i$ serve as elements. For a given ordered splitting, the root $\pmb{{\alpha}}_0$ corresponds to the kinematic factor
\bea
(-)^{|\pmb{\alpha}_0|}(\epsilon_1\cdot  f_{\alpha^0_2}\cdot f_{\alpha^0_3}\cdots  f_{\alpha^0_{|0|-1}}\cdot\epsilon_n)\,,
\eea
wile other ordered sets $\pmb{{\alpha}}_i$ with $i\neq 0$ correspond to
\bea
\epsilon_{R_i}\cdot  f_{\alpha^i_{|i|-1}}\cdots  f_{\alpha^i_2}\cdot  f_{\alpha^i_1}\cdot Z_{\alpha^i_1}\,.
\eea
In the above factors, the tensor $f_i^{\mu\nu}$ is defined by
\bea
f_i^{\mu\nu}\equiv k_i^\mu\epsilon_i^\nu-\epsilon_i^\mu k_i^\nu\,.
\eea
The combinatory momentum $Z_{\alpha^i_1}$ is the sum of momenta of external legs satisfying two conditions: (1) legs
at the LHS of $\alpha^i_1$ in the color-ordering, (2) legs belong to $\pmb{{\alpha}}_j$ at the LHS of $\pmb{{\alpha}}_i$
in the ordered splitting, i.e., $j<i$. The coefficient $C^\epsilon(\sigma)$
is the sum of contributions from all correct ordered splittings. Analogous algorithm holds for evaluating $C^{\W\epsilon}(\sigma')$.

To illustrate the procedure more clearly, let us consider the $4$-point BCJ numerator $C^\epsilon(32)$ for color-ordering $1\dot{<}3\dot{<}2\dot{<}4$. The reference ordering is chosen to be $4\prec3\prec2\prec1$. The roots $\pmb{{\alpha}}_0$ have following candidates: $\{1,4\}$, $\{1,2,4\}$, $\{1,3,4\}$, $\{1,3,2,4\}$. The ordered set $\{1,2,3,4\}$ violates the color-ordering $3\dot{<}2$ therefore can be excluded. For the root $\pmb{{\alpha}}_0=\{1,4\}$, the lowest element in the reduced reference ordering $3\prec2$ is $3$, then one can construct ${\pmb{\alpha}}_1=\{3\}$ or $\pmb{{\alpha}}_1=\{2,3\}$. However, the ordered set $\{2,3\}$ violates the color-ordering $3\dot{<}2$ therefore is forbidden. Thus, we obtain the ordered splitting $\{\{1,4\},\{3\},\{2\}\}$ for the root $\{1,4\}$. Similarly, one can get $\{\{1,2,4\},\{3\}\}$, $\{\{1,3,4\},\{2\}\}$ and $\{\{1,3,2,4\}\}$ for other roots. After giving kinematic factors for these each splitting, the BCJ numerator $C^\epsilon(32)$ is found to be
\bea
(\epsilon_1\cdot\epsilon_4)(\epsilon_3\cdot Z_3)(\epsilon_2\cdot Z_2)-(\epsilon_1\cdot f_2\cdot \epsilon_4)(\epsilon_3\cdot Z_3)-(\epsilon_1\cdot f_3\cdot\epsilon_4)(\epsilon_2\cdot Z_2)+(\epsilon_1\cdot f_3\cdot f_2\cdot\epsilon_4)\,.
\eea

The coefficients $C^\epsilon(\sigma)$ and $C^{\W\epsilon}(\sigma')$, together with BAS amplitudes ${\cal A}_{\rm BAS}(1\sigma n|1\sigma'n)$
with two massive scalars,
provide the GR amplitude ${\cal A}^{\epsilon,\W\epsilon}_{\rm GR}(1,\cdots,n)$ with two massive gravitons, via the expansion in \eref{exp-GR}.

\subsection{Converting gravitons to scalars or photons}
\label{subsec-conver}

After obtaining the GR amplitude with two massive gravitons, one can convert gravitons to scalars or photons through the dimensional reduction procedure,
or applying differential operators to the GR amplitude, to get the desired SG and SPG amplitudes.

To get the SG amplitude with two massive scalars in $D$ dimensions, an effective way is to consider the GR amplitude with two massive gravitons
in $D+1$ dimensions. Roughly speaking, this method is to choose the polarization vectors of two massive gravitons to be in the extra dimension, while all other Lorentz vectors lie in $D$ dimensions \cite{Naculich:2015zha,Cachazo:2014xea}. More explicitly, one can set momenta and polarization vectors of external legs for the GR amplitude to be
\bea
& &P_1^\mu=(P_1^0,P_1^1,\cdots,P_1^{D-1}|0)\,,~~\epsilon_1^\mu=(0,0,\cdots,0|1)\,,~~\W\epsilon_1^\mu=(0,0,\cdots,0|1)\,,\nn
& &P_n^\mu=(P_n^0,P_n^1,\cdots,P_n^{D-1}|0)\,,~~\epsilon_n^\mu=(0,0,\cdots,0|1)\,,~~\W\epsilon_n^\mu=(0,0,\cdots,0|1)\,,\nn
& &k_a^\mu=(k_a^0,k_a^1,\cdots,k_a^{D-1}|0)\,,~~\epsilon_a^\mu=(\epsilon_a^0,\epsilon_a^1,\cdots,\epsilon_a^{D-1}|0)\,,
~~\W\epsilon_a^\mu=(\W\epsilon_a^0,\W\epsilon_a^1,\cdots,\W\epsilon_a^{D-1}|0)\,,
\eea
where $P_1$ and $P_n$ are massive momenta satisfying $P_1^2=P_n^2=m^2$, $k_a$ are massless momenta with $a\in\{2,\cdots,n-1\}$.
For each vector, components at the LHS of $|$ lie in $D$ dimensions, while the component at the RHS of $|$ lies in the extra dimension.
Under the above choices, two massive gravitons behave as two massive scalars in $D$ dimensions, thus the goal is achieved. This procedure is called the dimensional reduction. An equivalent approach
is to perform differential operators
\bea
{\cal T}^\epsilon[1n]\equiv{\partial\over\partial(\epsilon_1\cdot\epsilon_n)}\,,~~~~{\cal T}^{\W\epsilon}[1n]\equiv{\partial\over\partial(\W\epsilon_1\cdot\W\epsilon_n)}
\eea
to the GR amplitude, as proposed in \cite{Cheung:2017ems}, and proved in \cite{Zhou:2018wvn,Bollmann:2018edb}. To exclude contributions from the dilaton and $2$-form, one need to
turn all $\W\epsilon_i$ to $\epsilon_i$ at the final step.

\begin{figure}
  \centering
   \includegraphics[width=10cm]{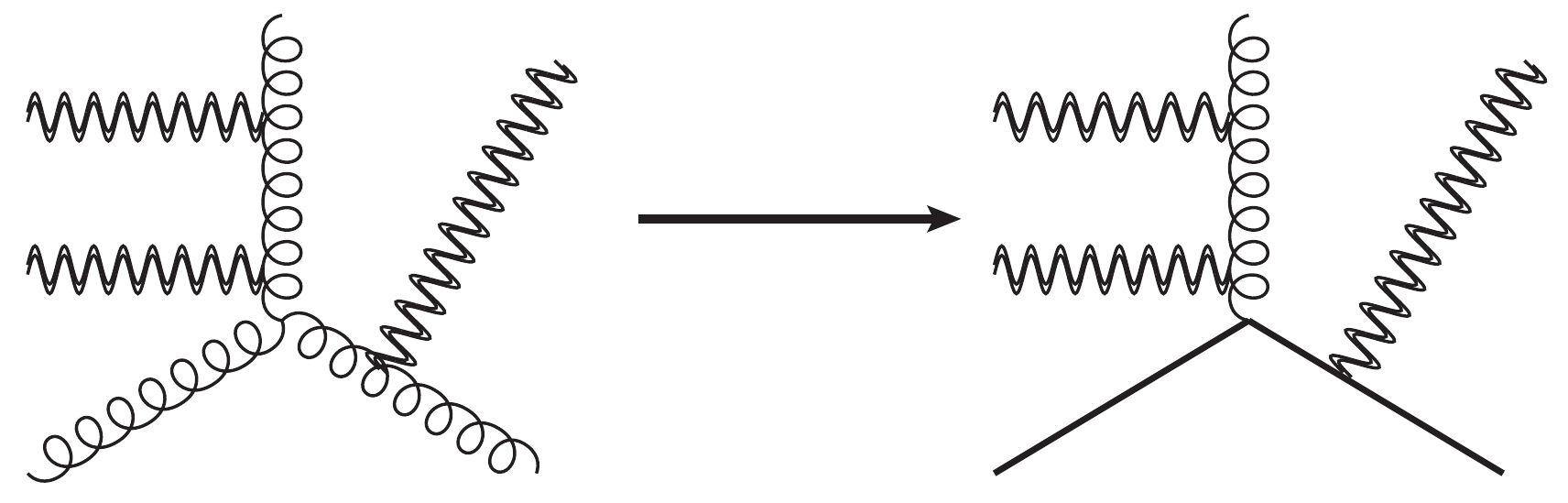} \\
  \caption{Converting two massive gluons to scalars when the number of gluons is $3$}\label{conver-g}
\end{figure}

To obtain the SPG amplitude with two massive scalars and one massless photon, we first apply the operator
\bea
{\cal T}^{\W\epsilon}[1bn]\equiv\Big({\partial\over\partial(P_1\cdot\W\epsilon_b)}-{\partial\over\partial(P_n\cdot\W\epsilon_b)}\Big){\cal T}^{\W\epsilon}[1n]
\eea
to the GR amplitude, to generate the single trace EYM amplitude ${\cal A}^{\epsilon,\W\epsilon}_{\rm EYM}(1bn;\{1,\cdots,n\}\setminus\{1,b,n\})$,
with two massive gluons $1\,,\,n$ and one massless gluon $b$, as well as $n-3$ gravitons in the set $\{1,\cdots,n\}\setminus\{1,b,n\}$. Then we use the operator ${\cal T}^\epsilon[1n]$ to convert massive gluons $1\,,\,n$ to scalars. After this manipulation, the remaining gluon $b$ can be identified as a photon. The reason is, after converting two massive gluons to scalars, the vector boson $b$ attached to the vertex including two scalars and $b$, bears the same structure with the photon-scalar vertex, as shown in Fig.\ref{conver-g}. When there is only one external gluon, the self interaction of gluons does not occur. Thus, from the angle of scattering amplitudes, a gluon can not be distinguished from a photon before adding the coupling constants. Consequently, we arrive at the SPG amplitude whose external particles are two scalars, one photon and $n-3$ gravitons\footnote{One may worry about the color-ordering $(1bn)$ from the single trace EYM amplitude, but for three elements, any way of changing the ordering leads to nothing but an overall sign, thus this color-ordering has no physical effect.}.

For the SPG amplitude with two massive scalars and one photon, it is hard to find a dimensional reduction manipulation which is equivalent to applying the differential operators, since the insertion operator
\bea
{\cal I}^{\W\epsilon}_{1bn}\equiv{\partial\over\partial(P_1\cdot\W\epsilon_b)}-{\partial\over\partial(P_n\cdot\W\epsilon_b)}
\eea
can not be interpreted directly by the dimensional reduction.

It is worth to emphasize that, the dimensional reduction technic, as well as the method of applying differential operators,
which are originally proposed for massless amplitudes, are also valid for amplitudes with two massive legs, as explained in \cite{Zhou:2020umm}.

\begin{figure}
  \centering
   \includegraphics[width=10cm]{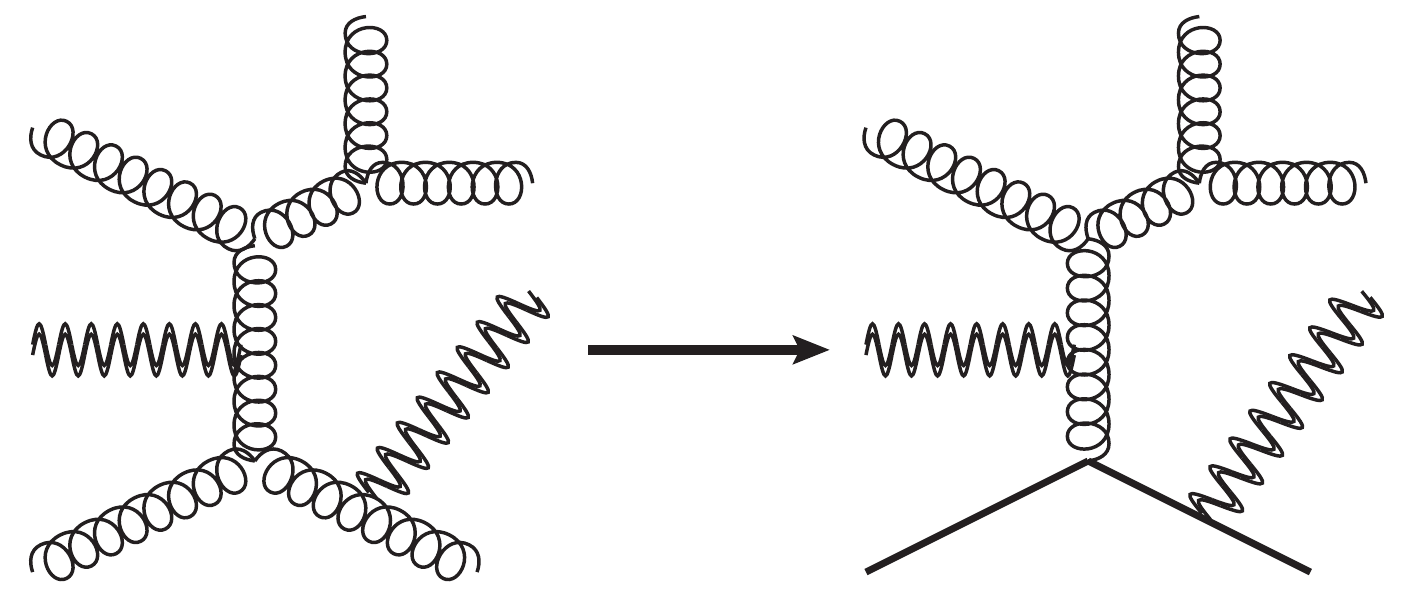} \\
  \caption{Converting two massive gluons to scalars when the number of gluons is larger than $3$}\label{conver-g2}
\end{figure}

\begin{figure}
  \centering
   \includegraphics[width=10cm]{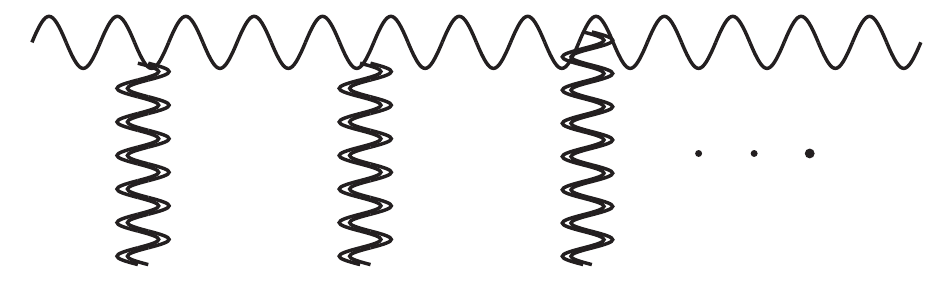} \\
  \caption{Photons interact with gravitons}\label{p-g}
\end{figure}

\begin{figure}
  \centering
   \includegraphics[width=10cm]{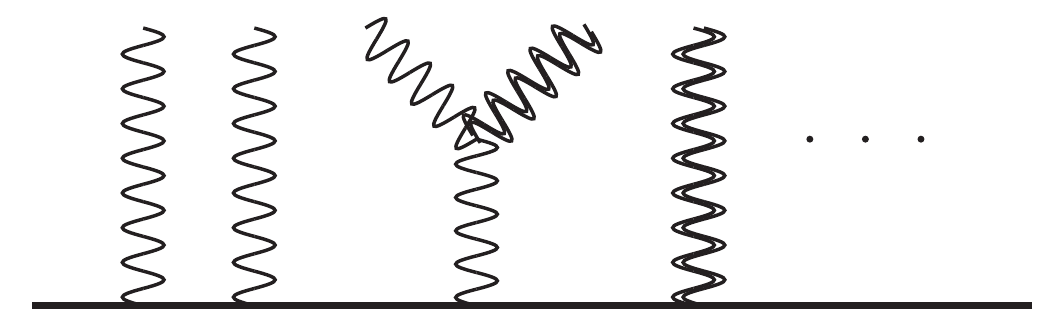} \\
  \caption{Scalars interact with photons}\label{s-p}
\end{figure}

Let us explain why it is hard to calculate the SPG amplitude with more than one photons by the method mentioned above. Suppose we naively apply this method, i.e., performing the trace operator ${\cal T}^{\W\epsilon}[1bc\cdots n]$ to the GR amplitude to create the single tace EYM amplitude with more gluons, the self interactions of gluons can not be removed by converting two gluons to scalars, as shown in Fig.\ref{conver-g2}. This fact indicates
that the remaining external gluons can never be identified as photons. But one can ask if there is other ways to turn gravitons to photons.
As well known, the photon-graviton amplitude can be generated from the GR amplitude by dimensional reduction \cite{Cachazo:2014xea}, or applying the operator \cite{Zhou:2018wvn}
\bea
{\cal T}^{\W\epsilon}_{X_{2m}}&\equiv&\sum_{\rho\in{\rm pair}}\,\prod_{i_k,j_k\in\rho}\,{\cal T}^{\W\epsilon}[i_k,j_k]\,.~~~~\label{h-p}
\eea
Here the set
\bea
\rho=\{(i_1,j_1),\cdots,(i_m,j_m)\}
\eea
is a partition of the length-$2m$ set including two massive scalars and all photons into pairs, with $i_1<i_2<\cdots<i_m$ and $i_t<j_t$, $\forall\,t$.
The summation is over all partitions $\rho$. For the case under consideration in this paper, one can not use the above manipulation
to convert gravitons to photons. The reason is, using the above method, the contributions from vertices that two photons interact with one graviton can not be avoid, as shown in Fig.\ref{p-g}, then two problems arise. First, for our purpose, the only required interaction which contains photons is that two massive scalars interact with one photon, as shown in Fig.\ref{s-p}. Secondly, for two types of vertices which include photons, the coupling constants are different. Since our method only concern on external states of amplitudes, how to separate different pieces with different coupling constants from a full amplitude becomes a hard obstacle. This is why we do not employ the dimensional reduction or the operator ${\cal T}^{\W\epsilon}_{X_{2m}}$ to convert gravitons to photons\footnote{In practical, if the number of photons is odd, there is another problem for the dimensional reduction method, or equivalently applying ${\cal T}^{\W\epsilon}_{X_{2m}}$. When number of photons is odd, at least one polarization vector $\epsilon_b^\mu$ lies in the extra dimension will contract with a vector lies in $D$ dimensions, thus the result will be $0$.}.

Before ending this part, we point out there are other alternative choices of expansions. For example, one can calculate the EYM amplitude
with two massive and one massless gluons by expanding it to BAS amplitudes, then convert two massive gluons to scalars, to get the SPG
amplitude with two massive scalars, one photon and $n-3$ gravitons. The algorithm for evaluating coefficients for the EYM amplitude can be
generated from the algorithm for computing coefficients for the GR amplitude via the differential operator ${\cal T}^{\W\epsilon}[1bn]$, as can be seen in \cite{Zhou:2019mbe}.
Thus, two methods are totally equivalent, since the differential operators only act on coefficients rather than basis in the expanded formula of GR amplitude \eref{exp-GR}.

\section{Scalar-graviton examples}
\label{secSG}

In this section, we use the method introduced in the previous section, to calculate SG amplitudes with two massive scalars and $n-2$ gravitons.
Before going to examples, let us do a little simplification of the algorithm. The dimensional reduction manipulation, as well as the differential operators, only act on two coefficients in the expanded GR amplitude in \eref{exp-GR}. The effect of them is turning all $\epsilon_1\cdot\epsilon_n$
and $\W\epsilon_1\cdot\W\epsilon_n$ to $1$, while annihilating all other terms do not contain both $\epsilon_1\cdot\epsilon_n$
and $\W\epsilon_1\cdot\W\epsilon_n$. Thus, one only need to consider ordered splittings with the root $\{1,n\}$, and turn
the corresponding terms in $C^\epsilon(\sigma)$ and $C^{\W\epsilon}(\sigma')$ to ${\cal C}^\epsilon(\sigma)$ and ${\cal C}^{\W\epsilon}(\sigma')$, by setting $\epsilon_1\cdot\epsilon_n=1$, $\W\epsilon_1\cdot\W\epsilon_n=1$,
respectively. Then we arrive at
\bea
{\cal \W A}_{\rm SG}({\bf1}_\varphi,2_h,\cdots,(n-1)_h,{\bf n}_\varphi)=\sum_\sigma\,\sum_{\sigma'}\,{\cal C}^\epsilon(\sigma){\cal A}_{\rm BAS}({\bf1}\sigma {\bf n}|{\bf1}\sigma'{\bf n}){\cal C}^{\W\epsilon}(\sigma')\,.~~~~\label{exp-SG-d2}
\eea
In this and next sections, we use the bold number to denote massive external particles.
The notation ${\cal \W A}_{\rm SG}$ stands for the SG amplitude which includes contributions from the dilaton and $2$-form.
The last step is turning all $\W\epsilon_i$ to $\epsilon_i$, to obtain the result
\bea
{\cal A}_{\rm SG}({\bf1}_\varphi,2_h,\cdots,(n-1)_h,{\bf n}_\varphi)=\sum_\sigma\,\sum_{\sigma'}\,{\cal C}^\epsilon(\sigma){\cal A}_{\rm BAS}({\bf1}\sigma {\bf n}|{\bf1}\sigma'{\bf n}){\cal C}^{\epsilon}(\sigma')\,.~~~~\label{exp-SG}
\eea
%

\subsection{$4$-point amplitude ${\cal A}_{\rm SG}({\bf 1}_\varphi,2_h,3_h,{\bf 4}_\varphi)$}

The simplest example is the $4$-point amplitude ${\cal A}_{\rm SG}({\bf1}_\varphi,2_h,3_h,{\bf4}_\varphi)$ with two massive scalars ${\bf1}_\varphi$, ${\bf4}_\varphi$, and two massless gravitons $2_h$, $3_h$.

According to the expansion in \eref{exp-SG}, we need to calculate the $4$-point BAS amplitudes
${\cal A}_{\rm BAS}({\bf1}\sigma {\bf4}|{\bf1}\sigma'{\bf4})$, as well as the coefficients ${\cal C}^\epsilon(\sigma)$ and ${\cal C}^{\W\epsilon}(\sigma')$.
Using the diagrammatic technic introduced in \S\ref{secalgo}, the $4$-point BAS amplitudes with two massive scalars ${\bf 1}$ and ${\bf 4}$
fixed at two ends in the color-orderings can be calculated as
\bea
{\cal A}_{\rm BAS}({\bf1}23{\bf4}|{\bf1}23{\bf4})&=&-{1\over s_{12}}-{1\over s_{23}}\,,\nn
{\cal A}_{\rm BAS}({\bf1}23{\bf4}|{\bf1}32{\bf4})&=&{\cal A}_{\rm BAS}({\bf1}32{\bf4}|{\bf1}23{\bf4})={1\over s_{23}}\,,\nn
{\cal A}_{\rm BAS}({\bf1}32{\bf4}|{\bf1}32{\bf4})&=&-{1\over s_{13}}-{1\over s_{23}}\,.\nn~~~~\label{4p-bas}
\eea
The coefficients can be computed by finding all correct ordered splittings, as discussed in \S\ref{secalgo}. The expressions for coefficients are given as
\bea
{\cal C}^\epsilon(23)&=&(\epsilon_3\cdot P_1)(\epsilon_2\cdot P_1)+\epsilon_3\cdot f_2\cdot P_1\,,\nn
{\cal C}^{\epsilon}(32)&=&(\epsilon_3\cdot P_1)(\epsilon_2\cdot P_{13})\,,~~~~\label{4p-coe}
\eea
with the reference ordering $4\prec3\prec2\prec1$.
Substituting these ingredients into the expansion \eref{exp-SG-d2}, we get the desired SG amplitude expressed as
\bea
{\cal A}_{\rm SG}({\bf1}_\varphi,2_h,3_h,{\bf4}_\varphi)&=&-\Big((\epsilon_3\cdot P_1)(\epsilon_2\cdot P_1)+\epsilon_3\cdot f_2\cdot P_1\Big)^2
\Big({1\over s_{12}}+{1\over s_{23}}\Big)\nn
& &+
{2\Big((\epsilon_3\cdot P_1)(\epsilon_2\cdot P_1)+\epsilon_3\cdot f_2\cdot P_1\Big)\Big((\epsilon_3\cdot P_1)(\epsilon_2\cdot P_{13})\Big)\over s_{23}}\nn
& &-\Big((\epsilon_3\cdot P_1)(\epsilon_2\cdot P_{13})\Big)^2\Big({1\over s_{13}}+{1\over s_{23}}\Big)\,.
\eea
It is straightforward to verify the equivalence between this expression and the formula obtained in \cite{Bjerrum-Bohr:2019nws}.

\subsection{$5$-point amplitude ${\cal A}_{\rm SG}({\bf 1}_\varphi,2_h,3_h,4_h,{\bf 5}_\varphi)$}

Then we consider the $5$-point example ${\cal A}_{\rm SG}({\bf 1}_\varphi,2_h,3_h,4_h,{\bf 5}_\varphi)$, with two massive scalars ${\bf1}_\varphi$, ${\bf5}_\varphi$, and three massless gravitons $2_h$, $3_h$, $4_h$. The
$5$-point BAS amplitudes with two massive legs ${\bf1}$, ${\bf5}$ and color-orderings $({\bf1}234{\bf5}|{\bf1}\sigma{\bf5})$ can be computed as
\bea
{\cal A}_{\rm BAS}({\bf1}234{\bf5}|{\bf1}234{\bf5})&=&{1\over s_{234}s_{34}}+{1\over s_{12}s_{34}}+{1\over s_{12}s_{45}}+{1\over s_{23}s_{45}}+{1\over s_{23}s_{234}}\,,\nn
{\cal A}_{\rm BAS}({\bf1}234{\bf5}|{\bf1}243{\bf5})&=&-{1\over s_{34}s_{234}}-{1\over s_{12}s_{34}}\,,\nn
{\cal A}_{\rm BAS}({\bf1}234{\bf5}|{\bf1}324{\bf5})&=&-{1\over s_{23}s_{45}}-{1\over s_{23}s_{234}}\,,\nn
{\cal A}_{\rm BAS}({\bf1}234{\bf5}|{\bf1}342{\bf5})&=&-{1\over s_{34}s_{234}}\,,\nn
{\cal A}_{\rm BAS}({\bf1}234{\bf5}|{\bf1}423{\bf5})&=&-{1\over s_{23}s_{234}}\,,\nn
{\cal A}_{\rm BAS}({\bf1}234{\bf5}|{\bf1}432{\bf5})&=&{1\over s_{234}s_{34}}+{1\over s_{234}s_{23}}\,.~~~~\label{5p-bas}
\eea
Other $5$-point BAS amplitudes can be obtained from them by changing nodes, for example, ${\cal A}_{\rm BAS}({\bf1}324{\bf5}|{\bf1}243{\bf5})$
can be generated from ${\cal A}_{\rm BAS}({\bf1}234{\bf5}|{\bf1}342{\bf5})$ via the replacement $2\to3\,,\,3\to2$.
The coefficients ${\cal C}^\epsilon(\sigma)$
can be calculated as
\bea
{\cal C}^\epsilon(234)&=&(\epsilon_4\cdot P_1)(\epsilon_3\cdot P_1)(\epsilon_2\cdot P_1)+(\epsilon_4\cdot P_1)(\epsilon_3\cdot f_2\cdot P_1)+(\epsilon_4\cdot f_3\cdot P_1)(\epsilon_2\cdot P_1)\nn
& &+(\epsilon_4\cdot f_2\cdot P_1)(\epsilon_3\cdot P_{12})+(\epsilon_4\cdot f_3\cdot f_2\cdot P_1)\,,\nn
{\cal C}^\epsilon(243)&=&(\epsilon_4\cdot P_1)(\epsilon_3\cdot P_{14})(\epsilon_2\cdot P_1)+(\epsilon_4\cdot f_2\cdot P_1)(\epsilon_3\cdot P_{124})\,,\nn
{\cal C}^\epsilon(324)&=&(\epsilon_4\cdot P_1)(\epsilon_3\cdot P_1)(\epsilon_2\cdot P_{13})
+(\epsilon_4\cdot f_2\cdot P_1)(\epsilon_3\cdot P_1)+(\epsilon_4\cdot f_3\cdot P_1)(\epsilon_2\cdot P_{13})\nn
& &+(\epsilon_4\cdot f_2\cdot f_3\cdot P_1)\,,\nn
{\cal C}^\epsilon(342)&=&(\epsilon_4\cdot P_1)(\epsilon_3\cdot P_1)(\epsilon_2\cdot P_{134})
+(\epsilon_4\cdot f_3\cdot P_1)(\epsilon_2\cdot P_{134})\,,\nn
{\cal C}^\epsilon(423)&=&(\epsilon_4\cdot P_1)(\epsilon_3\cdot P_{14})(\epsilon_2\cdot P_{14})
+(\epsilon_4\cdot P_1)(\epsilon_3\cdot f_2\cdot P_{14})\,,\nn
{\cal C}^\epsilon(432)&=&(\epsilon_4\cdot P_1)(\epsilon_3\cdot P_{14})(\epsilon_2\cdot P_{134})\,,~~~~\label{5p-coe}
\eea
with the reference ordering $5\prec4\prec3\prec2\prec1$.
Substituting \eref{5p-bas} and \eref{5p-coe} into \eref{exp-SG},
we arrive at the SG amplitude with two massive scalars, as exhibited in Appendix \ref{secappen1}.

Higher points SG amplitudes with two massive scalars follow by the same method analogously. Although the lengths of expressions for higher points
amplitudes are long, the calculation can be easily realized in MATHEMATIC.

\section{Scalar-photon-graviton examples}
\label{secSPG}

In this section we consider SPG amplitudes with two massive scalars, one massless photon and $n-3$ massless gravitons. As discussed in \S\ref{secalgo},
such amplitudes can be obtained by applying ${\cal T}^\epsilon[1n]$ and ${\cal T}^{\W\epsilon}[1bn]\equiv{\cal I}^{\W\epsilon}_{1bn}{\cal T}^{\W\epsilon}[1n]$ to the GR amplitude, then turning all $\W\epsilon_i$ to $\epsilon_i$. Performing ${\cal T}^\epsilon[1n]$ and ${\cal T}^{\W\epsilon}[1n]$ to the GR amplitude gives rise to the SG amplitude ${\cal \W A}_{\rm SG}({\bf1}_\varphi,2_h,\cdots,(n-1)_h,{\bf n}_\varphi)$ discussed in the previous section, thus a more
efficient way is to apply the insertion operator ${\cal I}^{\W\epsilon}_{1bn}$ to ${\cal \W A}_{\rm SG}({\bf1}_\varphi,2_h,\cdots,(n-1)_h,{\bf n}_\varphi)$, then turn all $\W\epsilon_i$ to $\epsilon_i$.

\subsection{$4$-point amplitude ${\cal A}_{\rm SPG}({\bf 1}_\varphi,2_p,3_h,{\bf 4}_\varphi)$}

Our first example is the $4$-point amplitude ${\cal A}_{\rm SPG}({\bf 1}_\varphi,2_p,3_h,{\bf 4}_\varphi)$, with two massive scalars ${\bf1}_\varphi$, ${\bf4}_\varphi$, one massless photon $2_p$, and one massless graviton $3_h$. As discussed above, we will apply the operator
\bea
{\cal I}^{\W\epsilon}_{124}\equiv{\partial\over\partial(\W\epsilon_2\cdot P_1)}-{\partial\over\partial(\W\epsilon_2\cdot k_4)}\,,
\eea
to the amplitude ${\cal \W A}_{\rm SG}({\bf 1}_\varphi,2_h,3_h,{\bf 4}_\varphi)$.
Using the BAS amplitudes in \eref{4p-bas} and coefficients in \eref{4p-coe} (coefficients ${\cal C}^{\W\epsilon}(\sigma)$ can be generated from ${\cal C}^{\epsilon}(\sigma)$ by turning all $\epsilon_i$ to $\W\epsilon_i$), we have
\bea
{\cal \W A}_{\rm SG}({\bf1}_\varphi,2_h,3_h,{\bf4}_\varphi)&=&-\Big((\epsilon_3\cdot P_1)(\epsilon_2\cdot P_1)+\epsilon_3\cdot f_2\cdot P_1\Big)
\Big({1\over s_{12}}+{1\over s_{23}}\Big)\Big((\W\epsilon_3\cdot P_1)(\W\epsilon_2\cdot P_1)+\W\epsilon_3\cdot \W f_2\cdot P_1\Big)\nn
& &+
{\Big((\epsilon_3\cdot P_1)(\epsilon_2\cdot P_1)+\epsilon_3\cdot f_2\cdot P_1\Big)\Big((\W\epsilon_3\cdot P_1)(\W\epsilon_2\cdot P_{13})\Big)\over s_{23}}\nn
& &+
{\Big((\epsilon_3\cdot P_1)(\epsilon_2\cdot P_{13})\Big)\Big((\W\epsilon_3\cdot P_1)(\W\epsilon_2\cdot P_1)+\W\epsilon_3\cdot\W f_2\cdot P_1\Big)\over s_{23}}\nn
& &-\Big((\epsilon_3\cdot P_1)(\epsilon_2\cdot P_{13})\Big)\Big({1\over s_{13}}+{1\over s_{23}}\Big)\Big((\W\epsilon_3\cdot P_1)(\W\epsilon_2\cdot P_{13})\Big)\,.
\eea
After applying ${\cal I}^{\W\epsilon}_{124}$, and turning all $\W\epsilon_i$ to $\epsilon_i$,
we get the result
\bea
& &{\cal A}_{\rm SPG}({\bf1}_\varphi,2_p,3_h,{\bf4}_\varphi)\nn
&=&-\Big((\epsilon_3\cdot P_1)(\epsilon_2\cdot P_1)+\epsilon_3\cdot f_2\cdot P_1\Big)
\Big({1\over s_{12}}+{1\over s_{23}}\Big)\Big((\epsilon_3\cdot P_1)+\epsilon_3\cdot k_2\Big)\nn
& &+
{\Big((\epsilon_3\cdot P_1)+\epsilon_3\cdot k_2\Big)\Big((\epsilon_3\cdot P_1)(\epsilon_2\cdot P_{13})\Big)+\Big((\epsilon_3\cdot P_1)(\epsilon_2\cdot P_1)+\epsilon_3\cdot f_2\cdot P_1\Big)(\epsilon_3\cdot P_1)\over s_{23}}\nn
& &-\Big((\epsilon_3\cdot P_1)(\epsilon_2\cdot P_{13})\Big)\Big({1\over s_{13}}+{1\over s_{23}}\Big)(\epsilon_3\cdot P_1)\,.
\eea
%

\subsection{$5$-point amplitude ${\cal A}_{\rm SPG}({\bf 1}_\varphi,2_p,3_h,4_h,{\bf 5}_\varphi)$}

The next example is the $5$-point amplitude ${\cal A}_{\rm SPG}({\bf 1}_\varphi,2_p,3_h,4_h,{\bf 5}_\varphi)$,
with two massive scalars ${\bf 1}_\varphi$, ${\bf 5}_\varphi$, one massless photon $2_p$, and two massless gravitons $3_h$, $4_h$.
The computation of this amplitude follows the similar process.
Substituting BAS amplitudes in \eref{5p-bas} and coefficients in \eref{5p-coe} into \eref{exp-SG-d2}, one can get the SG amplitude
${\cal \W A}_{\rm SG}({\bf1}_\varphi, 2_h,3_h,4_h,{\bf5}_\varphi)$. Then the amplitude ${\cal A}_{\rm SPG}({\bf 1}_\varphi,2_p,3_h,4_h,{\bf 5}_\varphi)$ can be obtained by applying the insertion operator ${\cal I}^{\W\epsilon}_{125}$ to ${\cal \W A}_{\rm SG}({\bf1}_\varphi, 2_h,3_h,4_h,{\bf5}_\varphi)$ and turning all $\W\epsilon_i$ to $\epsilon_i$.
The full result of ${\cal A}_{\rm SPG}({\bf 1}_\varphi,2_p,3_h,4_h,{\bf 5}_\varphi)$ is shown in the Appendix \ref{secappen1}.
Higher points SPG amplitudes with two massive scalars and one photon can be calculated by the same method analogously.

\section*{Acknowledgments}

This work is supported by Chinese NSF funding under
contracts No.11805163, as well as NSF of Jiangsu Province under Grant No.BK20180897.

\appendix

\section{Full results for $5$-point examples}
\label{secappen1}

In this appendix, we show the full results of two $5$-point examples.

The SG amplitude ${\cal A}_{\rm SG}({\bf1}_\varphi,2_h,3_h,4_h,{\bf5}_\varphi)$ is given as follows,
\bea
{\cal A}_{\rm SG}({\bf1}_\varphi,2_h,3_h,4_h,{\bf5}_\varphi)={\bf A}_1+{\bf A}_2\,,
\eea
where
\bea
& &{\bf A}_1=\nn
& &\Big((\epsilon_4\cdot P_1)(\epsilon_3\cdot P_1)(\epsilon_2\cdot P_1)+(\epsilon_4\cdot P_1)(\epsilon_3\cdot f_2\cdot P_1)+(\epsilon_4\cdot f_3\cdot P_1)(\epsilon_2\cdot P_1)+(\epsilon_4\cdot f_2\cdot P_1)(\epsilon_3\cdot P_{12})\nn
& &+(\epsilon_4\cdot f_3\cdot f_2\cdot P_1)\Big)^2\Big({1\over s_{234}s_{34}}+{1\over s_{12}s_{34}}+{1\over s_{12}s_{45}}+{1\over s_{23}s_{45}}+{1\over s_{23}s_{234}}\Big)\nn
& &-2\Big((\epsilon_4\cdot P_1)(\epsilon_3\cdot P_1)(\epsilon_2\cdot P_1)+(\epsilon_4\cdot P_1)(\epsilon_3\cdot f_2\cdot P_1)+(\epsilon_4\cdot f_3\cdot P_1)(\epsilon_2\cdot P_1)+(\epsilon_4\cdot f_2\cdot P_1)(\epsilon_3\cdot P_{12})\nn
& &+(\epsilon_4\cdot f_3\cdot f_2\cdot P_1)\Big)\Big({1\over s_{34}s_{234}}+{1\over s_{12}s_{34}}\Big)
\Big((\epsilon_4\cdot P_1)(\epsilon_3\cdot P_{14})(\epsilon_2\cdot P_1)+(\epsilon_4\cdot f_2\cdot P_1)(\epsilon_3\cdot P_{124})\Big)\nn
& &-2\Big((\epsilon_4\cdot P_1)(\epsilon_3\cdot P_1)(\epsilon_2\cdot P_1)+(\epsilon_4\cdot P_1)(\epsilon_3\cdot f_2\cdot P_1)+(\epsilon_4\cdot f_3\cdot P_1)(\epsilon_2\cdot P_1)+(\epsilon_4\cdot f_2\cdot P_1)(\epsilon_3\cdot P_{12})\nn
& &+(\epsilon_4\cdot f_3\cdot f_2\cdot P_1)\Big)\Big({1\over s_{23}s_{45}}+{1\over s_{23}s_{234}}\Big)\Big((\epsilon_4\cdot P_1)(\epsilon_3\cdot P_1)(\epsilon_2\cdot P_{13})
+(\epsilon_4\cdot f_2\cdot P_1)(\epsilon_3\cdot P_1)\nn
& &+(\epsilon_4\cdot f_3\cdot P_1)(\epsilon_2\cdot P_{13})+(\epsilon_4\cdot f_2\cdot f_3\cdot P_1)\Big)\nn
& &-2\Big((\epsilon_4\cdot P_1)(\epsilon_3\cdot P_1)(\epsilon_2\cdot P_1)+(\epsilon_4\cdot P_1)(\epsilon_3\cdot f_2\cdot P_1)+(\epsilon_4\cdot f_3\cdot P_1)(\epsilon_2\cdot P_1)+(\epsilon_4\cdot f_2\cdot P_1)(\epsilon_3\cdot P_{12})\nn
& &+(\epsilon_4\cdot f_3\cdot f_2\cdot P_1)\Big){1\over s_{34}s_{234}}\Big((\epsilon_4\cdot P_1)(\epsilon_3\cdot P_1)(\epsilon_2\cdot P_{134})
+(\epsilon_4\cdot f_3\cdot P_1)(\epsilon_2\cdot P_{134})\Big)\nn
& &-2\Big((\epsilon_4\cdot P_1)(\epsilon_3\cdot P_1)(\epsilon_2\cdot P_1)+(\epsilon_4\cdot P_1)(\epsilon_3\cdot f_2\cdot P_1)+(\epsilon_4\cdot f_3\cdot P_1)(\epsilon_2\cdot P_1)+(\epsilon_4\cdot f_2\cdot P_1)(\epsilon_3\cdot P_{12})\nn
& &+(\epsilon_4\cdot f_3\cdot f_2\cdot P_1)\Big){1\over s_{23}s_{234}}\Big((\epsilon_4\cdot P_1)(\epsilon_3\cdot P_{14})(\epsilon_2\cdot P_{14})
+(\epsilon_4\cdot P_1)(\epsilon_3\cdot f_2\cdot P_{14})\Big)\nn
& &+2\Big((\epsilon_4\cdot P_1)(\epsilon_3\cdot P_1)(\epsilon_2\cdot P_1)+(\epsilon_4\cdot P_1)(\epsilon_3\cdot f_2\cdot P_1)+(\epsilon_4\cdot f_3\cdot P_1)(\epsilon_2\cdot P_1)+(\epsilon_4\cdot f_2\cdot P_1)(\epsilon_3\cdot P_{12})\nn
& &+(\epsilon_4\cdot f_3\cdot f_2\cdot P_1)\Big)\Big({1\over s_{34}s_{234}}+{1\over s_{23}s_{234}}\Big)\Big((\epsilon_4\cdot P_1)(\epsilon_3\cdot P_{14})(\epsilon_2\cdot P_{134})\Big)\nn
& &+\Big((\epsilon_4\cdot P_1)(\epsilon_3\cdot P_{14})(\epsilon_2\cdot P_1)+(\epsilon_4\cdot f_2\cdot P_1)(\epsilon_3\cdot P_{124})\Big)^2
\Big({1\over s_{234}s_{34}}+{1\over s_{12}s_{34}}+{1\over s_{12}s_{35}}+{1\over s_{24}s_{35}}\nn
& &+{1\over s_{24}s_{234}}\Big)-2\Big((\epsilon_4\cdot P_1)(\epsilon_3\cdot P_{14})(\epsilon_2\cdot P_1)+(\epsilon_4\cdot f_2\cdot P_1)(\epsilon_3\cdot P_{124})\Big)
{1\over s_{24}s_{234}}\Big((\epsilon_4\cdot P_1)(\epsilon_3\cdot P_1)(\epsilon_2\cdot P_{13})\nn
& &+(\epsilon_4\cdot f_2\cdot P_1)(\epsilon_3\cdot P_1)+(\epsilon_4\cdot f_3\cdot P_1)(\epsilon_2\cdot P_{13})+(\epsilon_4\cdot f_2\cdot f_3\cdot P_1)\Big)\nn
& &+2\Big((\epsilon_4\cdot P_1)(\epsilon_3\cdot P_{14})(\epsilon_2\cdot P_1)+(\epsilon_4\cdot f_2\cdot P_1)(\epsilon_3\cdot P_{124})\Big)\Big({1\over s_{34}s_{234}}+{1\over s_{24}s_{234}}\Big)\nn
& &\times\Big((\epsilon_4\cdot P_1)(\epsilon_3\cdot P_1)(\epsilon_2\cdot P_{134})+(\epsilon_4\cdot f_3\cdot P_1)(\epsilon_2\cdot P_{134})\Big)\nn
& &-2\Big((\epsilon_4\cdot P_1)(\epsilon_3\cdot P_{14})(\epsilon_2\cdot P_1)+(\epsilon_4\cdot f_2\cdot P_1)(\epsilon_3\cdot P_{124})\Big)
\Big({1\over s_{23}s_{35}}+{1\over s_{24}s_{234}}\Big)\nn
& &\times\Big((\epsilon_4\cdot P_1)(\epsilon_3\cdot P_{14})(\epsilon_2\cdot P_{14})
+(\epsilon_4\cdot P_1)(\epsilon_3\cdot f_2\cdot P_{14})\Big)\nn
& &-2\Big((\epsilon_4\cdot P_1)(\epsilon_3\cdot P_{14})(\epsilon_2\cdot P_1)+(\epsilon_4\cdot f_2\cdot P_1)(\epsilon_3\cdot P_{124})\Big)
{1\over s_{34}s_{234}}\Big((\epsilon_4\cdot P_1)(\epsilon_3\cdot P_{14})(\epsilon_2\cdot P_{134})\Big)\,,
\eea
\bea
& &{\bf A}_2=\nn
& &\Big((\epsilon_4\cdot P_1)(\epsilon_3\cdot P_1)(\epsilon_2\cdot P_{13})
+(\epsilon_4\cdot f_2\cdot P_1)(\epsilon_3\cdot P_1)+(\epsilon_4\cdot f_3\cdot P_1)(\epsilon_2\cdot P_{13})+(\epsilon_4\cdot f_2\cdot f_3\cdot P_1)\Big)^2\nn
& &\times\Big({1\over s_{234}s_{24}}+{1\over s_{13}s_{24}}+{1\over s_{13}s_{45}}+{1\over s_{23}s_{45}}+{1\over s_{23}s_{234}}\Big)\nn
& &-2\Big((\epsilon_4\cdot P_1)(\epsilon_3\cdot P_1)(\epsilon_2\cdot P_{13})
+(\epsilon_4\cdot f_2\cdot P_1)(\epsilon_3\cdot P_1)+(\epsilon_4\cdot f_3\cdot P_1)(\epsilon_2\cdot P_{13})+(\epsilon_4\cdot f_2\cdot f_3\cdot P_1)\Big)\nn
& &\times\Big({1\over s_{24}s_{234}}+{1\over s_{13}s_{24}}\Big)\Big((\epsilon_4\cdot P_1)(\epsilon_3\cdot P_1)(\epsilon_2\cdot P_{134})
+(\epsilon_4\cdot f_3\cdot P_1)(\epsilon_2\cdot P_{134})\Big)\nn
& &+2\Big((\epsilon_4\cdot P_1)(\epsilon_3\cdot P_1)(\epsilon_2\cdot P_{13})
+(\epsilon_4\cdot f_2\cdot P_1)(\epsilon_3\cdot P_1)+(\epsilon_4\cdot f_3\cdot P_1)(\epsilon_2\cdot P_{13})+(\epsilon_4\cdot f_2\cdot f_3\cdot P_1)\Big)\nn
& &\times\Big({1\over s_{24}s_{234}}+{1\over s_{23}s_{234}}\Big)\Big((\epsilon_4\cdot P_1)(\epsilon_3\cdot P_{14})(\epsilon_2\cdot P_{14})
+(\epsilon_4\cdot P_1)(\epsilon_3\cdot f_2\cdot P_{14})\Big)\nn
& &-2\Big((\epsilon_4\cdot P_1)(\epsilon_3\cdot P_1)(\epsilon_2\cdot P_{13})
+(\epsilon_4\cdot f_2\cdot P_1)(\epsilon_3\cdot P_1)+(\epsilon_4\cdot f_3\cdot P_1)(\epsilon_2\cdot P_{13})+(\epsilon_4\cdot f_2\cdot f_3\cdot P_1)\Big)\nn
& &\times{1\over s_{23}s_{234}}\Big((\epsilon_4\cdot P_1)(\epsilon_3\cdot P_{14})(\epsilon_2\cdot P_{134})\Big)\nn
& &+\Big((\epsilon_4\cdot P_1)(\epsilon_3\cdot P_1)(\epsilon_2\cdot P_{134})
+(\epsilon_4\cdot f_3\cdot P_1)(\epsilon_2\cdot P_{134})\Big)^2\Big({1\over s_{234}s_{24}}+{1\over s_{13}s_{24}}+{1\over s_{13}s_{25}}
+{1\over s_{34}s_{25}}+{1\over s_{34}s_{234}}\Big)\nn
& &-2\Big((\epsilon_4\cdot P_1)(\epsilon_3\cdot P_1)(\epsilon_2\cdot P_{134})
+(\epsilon_4\cdot f_3\cdot P_1)(\epsilon_2\cdot P_{134})\Big){1\over s_{24}s_{234}}\nn
& &\times\Big((\epsilon_4\cdot P_1)(\epsilon_3\cdot P_{14})(\epsilon_2\cdot P_{14})
+(\epsilon_4\cdot P_1)(\epsilon_3\cdot f_2\cdot P_{14})\Big)\nn
& &-2\Big((\epsilon_4\cdot P_1)(\epsilon_3\cdot P_1)(\epsilon_2\cdot P_{134})
+(\epsilon_4\cdot f_3\cdot P_1)(\epsilon_2\cdot P_{134})\Big)\Big({1\over s_{34}s_{25}}+{1\over s_{34}s_{234}}\Big)\nn
& &\times\Big((\epsilon_4\cdot P_1)(\epsilon_3\cdot P_{14})(\epsilon_2\cdot P_{134})\Big)\nn
& &+\Big((\epsilon_4\cdot P_1)(\epsilon_3\cdot P_{14})(\epsilon_2\cdot P_{14})
+(\epsilon_4\cdot P_1)(\epsilon_3\cdot f_2\cdot P_{14})\Big)^2\Big({1\over s_{234}s_{23}}+{1\over s_{14}s_{23}}+{1\over s_{14}s_{35}}+{1\over s_{24}s_{35}}+{1\over s_{24}s_{234}}\Big)\nn
& &-2\Big((\epsilon_4\cdot P_1)(\epsilon_3\cdot P_{14})(\epsilon_2\cdot P_{14})
+(\epsilon_4\cdot P_1)(\epsilon_3\cdot f_2\cdot P_{14})\Big)\Big({1\over s_{23}s_{234}}+{1\over s_{14}s_{23}}\Big)
\Big((\epsilon_4\cdot P_1)(\epsilon_3\cdot P_{14})(\epsilon_2\cdot P_{134})\Big)\nn
& &+\Big((\epsilon_4\cdot P_1)(\epsilon_3\cdot P_{14})(\epsilon_2\cdot P_{134})\Big)^2\Big({1\over s_{234}s_{23}}+{1\over s_{14}s_{23}}+{1\over s_{14}s_{25}}+{1\over s_{34}s_{25}}+{1\over s_{34}s_{234}}\Big)
\eea

The SPG amplitude ${\cal A}_{\rm SPG}({\bf1}_\varphi,2_p,3_h,4_h,{\bf5}_\varphi)$ is given by
\bea
{\cal A}_{\rm SPG}({\bf1}_\varphi,2_p,3_h,4_h,{\bf5}_\varphi)={\bf B}_1+{\bf B}_2+{\bf B}_3\,,
\eea
where
\bea
& &{\bf B}_1=\nn
&=&\Big((\epsilon_4\cdot P_1)(\epsilon_3\cdot P_1)(\epsilon_2\cdot P_1)+(\epsilon_4\cdot P_1)(\epsilon_3\cdot f_2\cdot P_1)+(\epsilon_4\cdot f_3\cdot P_1)(\epsilon_2\cdot P_1)+(\epsilon_4\cdot f_2\cdot P_1)(\epsilon_3\cdot P_{12})\nn
& &+(\epsilon_4\cdot f_3\cdot f_2\cdot P_1)\Big)\Big({1\over s_{234}s_{34}}+{1\over s_{12}s_{34}}+{1\over s_{12}s_{45}}+{1\over s_{23}s_{45}}+{1\over s_{23}s_{234}}\Big)\nn
& &\times\Big((\epsilon_4\cdot P_1)(\epsilon_3\cdot P_1)+(\epsilon_4\cdot P_1)(\epsilon_3\cdot k_2)+(\epsilon_4\cdot f_3\cdot P_1)+(\epsilon_4\cdot k_2)(\epsilon_3\cdot P_{12})+(\epsilon_4\cdot f_3\cdot k_2)\Big)\nn
& &-\Big((\epsilon_4\cdot P_1)(\epsilon_3\cdot P_1)+(\epsilon_4\cdot P_1)(\epsilon_3\cdot k_2)+(\epsilon_4\cdot f_3\cdot P_1)+(\epsilon_4\cdot k_2)(\epsilon_3\cdot P_{12})+(\epsilon_4\cdot f_3\cdot k_2)\Big)\nn
& &\times\Big({1\over s_{34}s_{234}}+{1\over s_{12}s_{34}}\Big)
\Big((\epsilon_4\cdot P_1)(\epsilon_3\cdot P_{14})(\epsilon_2\cdot P_1)+(\epsilon_4\cdot f_2\cdot P_1)(\epsilon_3\cdot P_{124})\Big)\nn
& &-\Big((\epsilon_4\cdot P_1)(\epsilon_3\cdot P_1)(\epsilon_2\cdot P_1)+(\epsilon_4\cdot P_1)(\epsilon_3\cdot f_2\cdot P_1)+(\epsilon_4\cdot f_3\cdot P_1)(\epsilon_2\cdot P_1)+(\epsilon_4\cdot f_2\cdot P_1)(\epsilon_3\cdot P_{12})\nn
& &+(\epsilon_4\cdot f_3\cdot f_2\cdot P_1)\Big)\Big({1\over s_{34}s_{234}}+{1\over s_{12}s_{34}}\Big)
\Big((\epsilon_4\cdot P_1)(\epsilon_3\cdot P_{14})+(\epsilon_4\cdot k_2)(\epsilon_3\cdot P_{124})\Big)\nn
& &-\Big((\epsilon_4\cdot P_1)(\epsilon_3\cdot P_1)+(\epsilon_4\cdot P_1)(\epsilon_3\cdot k_2)+(\epsilon_4\cdot f_3\cdot P_1)+(\epsilon_4\cdot k_2)(\epsilon_3\cdot P_{12})+(\epsilon_4\cdot f_3\cdot k_2)\Big)\nn
& &\times\Big({1\over s_{23}s_{45}}+{1\over s_{23}s_{234}}\Big)\Big((\epsilon_4\cdot P_1)(\epsilon_3\cdot P_1)(\epsilon_2\cdot P_{13})
+(\epsilon_4\cdot f_2\cdot P_1)(\epsilon_3\cdot P_1)\nn
& &+(\epsilon_4\cdot f_3\cdot P_1)(\epsilon_2\cdot P_{13})+(\epsilon_4\cdot f_2\cdot f_3\cdot P_1)\Big)\nn
& &-\Big((\epsilon_4\cdot P_1)(\epsilon_3\cdot P_1)(\epsilon_2\cdot P_1)+(\epsilon_4\cdot P_1)(\epsilon_3\cdot f_2\cdot P_1)+(\epsilon_4\cdot f_3\cdot P_1)(\epsilon_2\cdot P_1)+(\epsilon_4\cdot f_2\cdot P_1)(\epsilon_3\cdot P_{12})\nn
& &+(\epsilon_4\cdot f_3\cdot f_2\cdot P_1)\Big)\Big({1\over s_{23}s_{45}}+{1\over s_{23}s_{234}}\Big)\Big((\epsilon_4\cdot P_1)(\epsilon_3\cdot P_1)
+(\epsilon_4\cdot k_2)(\epsilon_3\cdot P_1)+(\epsilon_4\cdot f_3\cdot P_1)\Big)\nn
& &-\Big((\epsilon_4\cdot P_1)(\epsilon_3\cdot P_1)+(\epsilon_4\cdot P_1)(\epsilon_3\cdot k_2)+(\epsilon_4\cdot f_3\cdot P_1)+(\epsilon_4\cdot k_2)(\epsilon_3\cdot P_{12})+(\epsilon_4\cdot f_3\cdot k_2)\Big)\nn
& &{1\over s_{34}s_{234}}\Big((\epsilon_4\cdot P_1)(\epsilon_3\cdot P_1)(\epsilon_2\cdot P_{134})
+(\epsilon_4\cdot f_3\cdot P_1)(\epsilon_2\cdot P_{134})\Big)\nn
& &-\Big((\epsilon_4\cdot P_1)(\epsilon_3\cdot P_1)(\epsilon_2\cdot P_1)+(\epsilon_4\cdot P_1)(\epsilon_3\cdot f_2\cdot P_1)+(\epsilon_4\cdot f_3\cdot P_1)(\epsilon_2\cdot P_1)+(\epsilon_4\cdot f_2\cdot P_1)(\epsilon_3\cdot P_{12})\nn
& &+(\epsilon_4\cdot f_3\cdot f_2\cdot P_1)\Big){1\over s_{34}s_{234}}\Big((\epsilon_4\cdot P_1)(\epsilon_3\cdot P_1)
+(\epsilon_4\cdot f_3\cdot P_1)\Big)\nn
& &-\Big((\epsilon_4\cdot P_1)(\epsilon_3\cdot P_1)+(\epsilon_4\cdot P_1)(\epsilon_3\cdot k_2)+(\epsilon_4\cdot f_3\cdot P_1)+(\epsilon_4\cdot k_2)(\epsilon_3\cdot P_{12})+(\epsilon_4\cdot f_3\cdot k_2)\Big)\nn
& &\times{1\over s_{23}s_{234}}\Big((\epsilon_4\cdot P_1)(\epsilon_3\cdot P_{14})(\epsilon_2\cdot P_{14})
+(\epsilon_4\cdot P_1)(\epsilon_3\cdot f_2\cdot P_{14})\Big)\nn
& &-\Big((\epsilon_4\cdot P_1)(\epsilon_3\cdot P_1)(\epsilon_2\cdot P_1)+(\epsilon_4\cdot P_1)(\epsilon_3\cdot f_2\cdot P_1)+(\epsilon_4\cdot f_3\cdot P_1)(\epsilon_2\cdot P_1)+(\epsilon_4\cdot f_2\cdot P_1)(\epsilon_3\cdot P_{12})\nn
& &+(\epsilon_4\cdot f_3\cdot f_2\cdot P_1)\Big){1\over s_{23}s_{234}}\Big((\epsilon_4\cdot P_1)(\epsilon_3\cdot P_{14})
+(\epsilon_4\cdot P_1)(\epsilon_3\cdot k_2)\Big)\nn
& &+\Big((\epsilon_4\cdot P_1)(\epsilon_3\cdot P_1)+(\epsilon_4\cdot P_1)(\epsilon_3\cdot k_2)+(\epsilon_4\cdot f_3\cdot P_1)+(\epsilon_4\cdot k_2)(\epsilon_3\cdot P_{12})+(\epsilon_4\cdot f_3\cdot k_2)\Big)\nn
& &\times\Big({1\over s_{34}s_{234}}+{1\over s_{23}s_{234}}\Big)\Big((\epsilon_4\cdot P_1)(\epsilon_3\cdot P_{14})(\epsilon_2\cdot P_{134})\Big)\nn
& &+\Big((\epsilon_4\cdot P_1)(\epsilon_3\cdot P_1)(\epsilon_2\cdot P_1)+(\epsilon_4\cdot P_1)(\epsilon_3\cdot f_2\cdot P_1)+(\epsilon_4\cdot f_3\cdot P_1)(\epsilon_2\cdot P_1)+(\epsilon_4\cdot f_2\cdot P_1)(\epsilon_3\cdot P_{12})\nn
& &+(\epsilon_4\cdot f_3\cdot f_2\cdot P_1)\Big)\Big({1\over s_{34}s_{234}}+{1\over s_{23}s_{234}}\Big)\Big((\epsilon_4\cdot P_1)(\epsilon_3\cdot P_{14})\Big)\,,
\eea
\bea
& &{\bf B}_2=\nn
& &\Big((\epsilon_4\cdot P_1)(\epsilon_3\cdot P_{14})(\epsilon_2\cdot P_1)+(\epsilon_4\cdot f_2\cdot P_1)(\epsilon_3\cdot P_{124})\Big)
\Big({1\over s_{234}s_{34}}+{1\over s_{12}s_{34}}+{1\over s_{12}s_{35}}+{1\over s_{24}s_{35}}\nn
& &+{1\over s_{24}s_{234}}\Big)\Big((\epsilon_4\cdot P_1)(\epsilon_3\cdot P_{14})+(\epsilon_4\cdot k_2)(\epsilon_3\cdot P_{124})\Big)\nn
& &-\Big((\epsilon_4\cdot P_1)(\epsilon_3\cdot P_{14})+(\epsilon_4\cdot k_2)(\epsilon_3\cdot P_{124})\Big)
{1\over s_{24}s_{234}}\Big((\epsilon_4\cdot P_1)(\epsilon_3\cdot P_1)(\epsilon_2\cdot P_{13})\nn
& &+(\epsilon_4\cdot f_2\cdot P_1)(\epsilon_3\cdot P_1)+(\epsilon_4\cdot f_3\cdot P_1)(\epsilon_2\cdot P_{13})+(\epsilon_4\cdot f_2\cdot f_3\cdot P_1)\Big)\nn
& &-\Big((\epsilon_4\cdot P_1)(\epsilon_3\cdot P_{14})(\epsilon_2\cdot P_1)+(\epsilon_4\cdot f_2\cdot P_1)(\epsilon_3\cdot P_{124})\Big)
{1\over s_{24}s_{234}}\Big((\epsilon_4\cdot P_1)(\epsilon_3\cdot P_1)\nn
& &+(\epsilon_4\cdot k_2)(\epsilon_3\cdot P_1)+(\epsilon_4\cdot f_3\cdot P_1)\Big)\nn
& &+\Big((\epsilon_4\cdot P_1)(\epsilon_3\cdot P_{14})+(\epsilon_4\cdot k_2)(\epsilon_3\cdot P_{124})\Big)\Big({1\over s_{34}s_{234}}+{1\over s_{24}s_{234}}\Big)\nn
& &\times\Big((\epsilon_4\cdot P_1)(\epsilon_3\cdot P_1)(\epsilon_2\cdot P_{134})+(\epsilon_4\cdot f_3\cdot P_1)(\epsilon_2\cdot P_{134})\Big)\nn
& &+\Big((\epsilon_4\cdot P_1)(\epsilon_3\cdot P_{14})(\epsilon_2\cdot P_1)+(\epsilon_4\cdot f_2\cdot P_1)(\epsilon_3\cdot P_{124})\Big)\Big({1\over s_{34}s_{234}}+{1\over s_{24}s_{234}}\Big)\nn
& &\times\Big((\epsilon_4\cdot P_1)(\epsilon_3\cdot P_1)+(\epsilon_4\cdot f_3\cdot P_1)\Big)\nn
& &-\Big((\epsilon_4\cdot P_1)(\epsilon_3\cdot P_{14})+(\epsilon_4\cdot k_2)(\epsilon_3\cdot P_{124})\Big)
\Big({1\over s_{23}s_{35}}+{1\over s_{24}s_{234}}\Big)\nn
& &\times\Big((\epsilon_4\cdot P_1)(\epsilon_3\cdot P_{14})(\epsilon_2\cdot P_{14})
+(\epsilon_4\cdot P_1)(\epsilon_3\cdot f_2\cdot P_{14})\Big)\nn
& &-\Big((\epsilon_4\cdot P_1)(\epsilon_3\cdot P_{14})(\epsilon_2\cdot P_1)+(\epsilon_4\cdot f_2\cdot P_1)(\epsilon_3\cdot P_{124})\Big)
\Big({1\over s_{23}s_{35}}+{1\over s_{24}s_{234}}\Big)\nn
& &\times\Big((\epsilon_4\cdot P_1)(\epsilon_3\cdot P_{14})
+(\epsilon_4\cdot P_1)(\epsilon_3\cdot k_2)\Big)\nn
& &-\Big((\epsilon_4\cdot P_1)(\epsilon_3\cdot P_{14})+(\epsilon_4\cdot k_2)(\epsilon_3\cdot P_{124})\Big)
{1\over s_{34}s_{234}}\Big((\epsilon_4\cdot P_1)(\epsilon_3\cdot P_{14})(\epsilon_2\cdot P_{134})\Big)\nn
& &-\Big((\epsilon_4\cdot P_1)(\epsilon_3\cdot P_{14})(\epsilon_2\cdot P_1)+(\epsilon_4\cdot f_2\cdot P_1)(\epsilon_3\cdot P_{124})\Big)
{1\over s_{34}s_{234}}\Big((\epsilon_4\cdot P_1)(\epsilon_3\cdot P_{14})\Big)\nn
& &+\Big((\epsilon_4\cdot P_1)(\epsilon_3\cdot P_{14})(\epsilon_2\cdot P_{14})
+(\epsilon_4\cdot P_1)(\epsilon_3\cdot f_2\cdot P_{14})\Big)\Big({1\over s_{234}s_{23}}+{1\over s_{14}s_{23}}+{1\over s_{14}s_{35}}+{1\over s_{24}s_{35}}+{1\over s_{24}s_{234}}\Big)\nn
& &\times\Big((\epsilon_4\cdot P_1)(\epsilon_3\cdot P_{14})
+(\epsilon_4\cdot P_1)(\epsilon_3\cdot k_2)\Big)\nn
& &-\Big((\epsilon_4\cdot P_1)(\epsilon_3\cdot P_{14})
+(\epsilon_4\cdot P_1)(\epsilon_3\cdot k_2)\Big)\Big({1\over s_{23}s_{234}}+{1\over s_{14}s_{23}}\Big)
\Big((\epsilon_4\cdot P_1)(\epsilon_3\cdot P_{14})(\epsilon_2\cdot P_{134})\Big)\nn
& &-\Big((\epsilon_4\cdot P_1)(\epsilon_3\cdot P_{14})(\epsilon_2\cdot P_{14})
+(\epsilon_4\cdot P_1)(\epsilon_3\cdot f_2\cdot P_{14})\Big)\Big({1\over s_{23}s_{234}}+{1\over s_{14}s_{23}}\Big)
\Big((\epsilon_4\cdot P_1)(\epsilon_3\cdot P_{14})\Big)\nn
& &+\Big((\epsilon_4\cdot P_1)(\epsilon_3\cdot P_{14})(\epsilon_2\cdot P_{134})\Big)\Big({1\over s_{234}s_{23}}+{1\over s_{14}s_{23}}+{1\over s_{14}s_{25}}+{1\over s_{34}s_{25}}+{1\over s_{34}s_{234}}\Big)\Big((\epsilon_4\cdot P_1)(\epsilon_3\cdot P_{14})\Big)\,,
\eea
\bea
& &{\bf B}_3=\nn
& &\Big((\epsilon_4\cdot P_1)(\epsilon_3\cdot P_1)(\epsilon_2\cdot P_{13})
+(\epsilon_4\cdot f_2\cdot P_1)(\epsilon_3\cdot P_1)+(\epsilon_4\cdot f_3\cdot P_1)(\epsilon_2\cdot P_{13})+(\epsilon_4\cdot f_2\cdot f_3\cdot P_1)\Big)\nn
& &\times\Big({1\over s_{234}s_{24}}+{1\over s_{13}s_{24}}+{1\over s_{13}s_{45}}+{1\over s_{23}s_{45}}+{1\over s_{23}s_{234}}\Big)\Big((\epsilon_4\cdot P_1)(\epsilon_3\cdot P_1)
+(\epsilon_4\cdot k_2)(\epsilon_3\cdot P_1)+(\epsilon_4\cdot f_3\cdot P_1)\Big)\nn
& &-\Big((\epsilon_4\cdot P_1)(\epsilon_3\cdot P_1)
+(\epsilon_4\cdot k_2)(\epsilon_3\cdot P_1)+(\epsilon_4\cdot f_3\cdot P_1)\Big)\nn
& &\times\Big({1\over s_{24}s_{234}}+{1\over s_{13}s_{24}}\Big)\Big((\epsilon_4\cdot P_1)(\epsilon_3\cdot P_1)(\epsilon_2\cdot P_{134})
+(\epsilon_4\cdot f_3\cdot P_1)(\epsilon_2\cdot P_{134})\Big)\nn
& &-\Big((\epsilon_4\cdot P_1)(\epsilon_3\cdot P_1)(\epsilon_2\cdot P_{13})
+(\epsilon_4\cdot f_2\cdot P_1)(\epsilon_3\cdot P_1)+(\epsilon_4\cdot f_3\cdot P_1)(\epsilon_2\cdot P_{13})+(\epsilon_4\cdot f_2\cdot f_3\cdot P_1)\Big)\nn
& &\times\Big({1\over s_{24}s_{234}}+{1\over s_{13}s_{24}}\Big)\Big((\epsilon_4\cdot P_1)(\epsilon_3\cdot P_1)
+(\epsilon_4\cdot f_3\cdot P_1)\Big)\nn
& &+\Big((\epsilon_4\cdot P_1)(\epsilon_3\cdot P_1)
+(\epsilon_4\cdot k_2)(\epsilon_3\cdot P_1)+(\epsilon_4\cdot f_3\cdot P_1)\Big)\nn
& &\times\Big({1\over s_{24}s_{234}}+{1\over s_{23}s_{234}}\Big)\Big((\epsilon_4\cdot P_1)(\epsilon_3\cdot P_{14})(\epsilon_2\cdot P_{14})
+(\epsilon_4\cdot P_1)(\epsilon_3\cdot f_2\cdot P_{14})\Big)\nn
& &+\Big((\epsilon_4\cdot P_1)(\epsilon_3\cdot P_1)(\epsilon_2\cdot P_{13})
+(\epsilon_4\cdot f_2\cdot P_1)(\epsilon_3\cdot P_1)+(\epsilon_4\cdot f_3\cdot P_1)(\epsilon_2\cdot P_{13})+(\epsilon_4\cdot f_2\cdot f_3\cdot P_1)\Big)\nn
& &\times\Big({1\over s_{24}s_{234}}+{1\over s_{23}s_{234}}\Big)\Big((\epsilon_4\cdot P_1)(\epsilon_3\cdot P_{14})
+(\epsilon_4\cdot P_1)(\epsilon_3\cdot k_2)\Big)\nn
& &-\Big((\epsilon_4\cdot P_1)(\epsilon_3\cdot P_1)
+(\epsilon_4\cdot k_2)(\epsilon_3\cdot P_1)+(\epsilon_4\cdot f_3\cdot P_1)\Big){1\over s_{23}s_{234}}\Big((\epsilon_4\cdot P_1)(\epsilon_3\cdot P_{14})(\epsilon_2\cdot P_{134})\Big)\nn
& &-\Big((\epsilon_4\cdot P_1)(\epsilon_3\cdot P_1)(\epsilon_2\cdot P_{13})
+(\epsilon_4\cdot f_2\cdot P_1)(\epsilon_3\cdot P_1)+(\epsilon_4\cdot f_3\cdot P_1)(\epsilon_2\cdot P_{13})+(\epsilon_4\cdot f_2\cdot f_3\cdot P_1)\Big)\nn
& &\times{1\over s_{23}s_{234}}\Big((\epsilon_4\cdot P_1)(\epsilon_3\cdot P_{14})\Big)\nn
& &+\Big((\epsilon_4\cdot P_1)(\epsilon_3\cdot P_1)(\epsilon_2\cdot P_{134})
+(\epsilon_4\cdot f_3\cdot P_1)(\epsilon_2\cdot P_{134})\Big)\Big({1\over s_{234}s_{24}}+{1\over s_{13}s_{24}}+{1\over s_{13}s_{25}}
+{1\over s_{34}s_{25}}+{1\over s_{34}s_{234}}\Big)\nn
& &\times\Big((\epsilon_4\cdot P_1)(\epsilon_3\cdot P_1)
+(\epsilon_4\cdot f_3\cdot P_1)\Big)\nn
& &-\Big((\epsilon_4\cdot P_1)(\epsilon_3\cdot P_1)
+(\epsilon_4\cdot f_3\cdot P_1)\Big){1\over s_{24}s_{234}}\Big((\epsilon_4\cdot P_1)(\epsilon_3\cdot P_{14})(\epsilon_2\cdot P_{14})
+(\epsilon_4\cdot P_1)(\epsilon_3\cdot f_2\cdot P_{14})\Big)\nn
& &-\Big((\epsilon_4\cdot P_1)(\epsilon_3\cdot P_1)(\epsilon_2\cdot P_{134})
+(\epsilon_4\cdot f_3\cdot P_1)(\epsilon_2\cdot P_{134})\Big){1\over s_{24}s_{234}}\Big((\epsilon_4\cdot P_1)(\epsilon_3\cdot P_{14})
+(\epsilon_4\cdot P_1)(\epsilon_3\cdot k_2)\Big)\nn
& &-\Big((\epsilon_4\cdot P_1)(\epsilon_3\cdot P_1)
+(\epsilon_4\cdot f_3\cdot P_1)\Big)\Big({1\over s_{34}s_{25}}+{1\over s_{34}s_{234}}\Big)\Big((\epsilon_4\cdot P_1)(\epsilon_3\cdot P_{14})(\epsilon_2\cdot P_{134})\Big)\nn
& &-\Big((\epsilon_4\cdot P_1)(\epsilon_3\cdot P_1)(\epsilon_2\cdot P_{134})
+(\epsilon_4\cdot f_3\cdot P_1)(\epsilon_2\cdot P_{134})\Big)\Big({1\over s_{34}s_{25}}+{1\over s_{34}s_{234}}\Big)\Big((\epsilon_4\cdot P_1)(\epsilon_3\cdot P_{14})\Big)\,,
\eea


\begin{thebibliography}{}


\bibitem{Abbott:2016blz}
B.~P.~Abbott \textit{et al.} [LIGO Scientific and Virgo],
``Observation of Gravitational Waves from a Binary Black Hole Merger,''
Phys. Rev. Lett. \textbf{116}, no.6, 061102 (2016)
doi:10.1103/PhysRevLett.116.061102
[arXiv:1602.03837 [gr-qc]].

\bibitem{TheLIGOScientific:2017qsa}
B.~P.~Abbott \textit{et al.} [LIGO Scientific and Virgo],
``GW170817: Observation of Gravitational Waves from a Binary Neutron Star Inspiral,''
Phys. Rev. Lett. \textbf{119}, no.16, 161101 (2017)
doi:10.1103/PhysRevLett.119.161101
[arXiv:1710.05832 [gr-qc]].


\bibitem{Cachazo:2017jef}
F.~Cachazo and A.~Guevara,
``Leading Singularities and Classical Gravitational Scattering,''
JHEP \textbf{02}, 181 (2020)
doi:10.1007/JHEP02(2020)181
[arXiv:1705.10262 [hep-th]].

\bibitem{Guevara:2017csg}
A.~Guevara,
``Holomorphic Classical Limit for Spin Effects in Gravitational and Electromagnetic Scattering,''
JHEP \textbf{04}, 033 (2019)
doi:10.1007/JHEP04(2019)033
[arXiv:1706.02314 [hep-th]].

\bibitem{Damour:2017zjx}
T.~Damour,
``High-energy gravitational scattering and the general relativistic two-body problem,''
Phys. Rev. D \textbf{97}, no.4, 044038 (2018)
doi:10.1103/PhysRevD.97.044038
[arXiv:1710.10599 [gr-qc]].

\bibitem{Bjerrum-Bohr:2018xdl}
N.~E.~J.~Bjerrum-Bohr, P.~H.~Damgaard, G.~Festuccia, L.~Plant and P.~Vanhove,
``General Relativity from Scattering Amplitudes,''
Phys. Rev. Lett. \textbf{121}, no.17, 171601 (2018)
doi:10.1103/PhysRevLett.121.171601
[arXiv:1806.04920 [hep-th]].

\bibitem{Levi:2018nxp}
M.~Levi,
``Effective Field Theories of Post-Newtonian Gravity: A comprehensive review,''
Rept. Prog. Phys. \textbf{83}, no.7, 075901 (2020)
doi:10.1088/1361-6633/ab12bc
[arXiv:1807.01699 [hep-th]].

\bibitem{Cheung:2018wkq}
C.~Cheung, I.~Z.~Rothstein and M.~P.~Solon,
``From Scattering Amplitudes to Classical Potentials in the Post-Minkowskian Expansion,''
Phys. Rev. Lett. \textbf{121}, no.25, 251101 (2018)
doi:10.1103/PhysRevLett.121.251101
[arXiv:1808.02489 [hep-th]].

\bibitem{Chung:2018kqs}
M.~Z.~Chung, Y.~T.~Huang, J.~W.~Kim and S.~Lee,
``The simplest massive S-matrix: from minimal coupling to Black Holes,''
JHEP \textbf{04}, 156 (2019)
doi:10.1007/JHEP04(2019)156
[arXiv:1812.08752 [hep-th]].

\bibitem{Bern:2019nnu}
Z.~Bern, C.~Cheung, R.~Roiban, C.~H.~Shen, M.~P.~Solon and M.~Zeng,
``Scattering Amplitudes and the Conservative Hamiltonian for Binary Systems at Third Post-Minkowskian Order,''
Phys. Rev. Lett. \textbf{122}, no.20, 201603 (2019)
doi:10.1103/PhysRevLett.122.201603
[arXiv:1901.04424 [hep-th]].

\bibitem{Bern:2019crd}
Z.~Bern, C.~Cheung, R.~Roiban, C.~H.~Shen, M.~P.~Solon and M.~Zeng,
``Black Hole Binary Dynamics from the Double Copy and Effective Theory,''
JHEP \textbf{10}, 206 (2019)
doi:10.1007/JHEP10(2019)206
[arXiv:1908.01493 [hep-th]].

\bibitem{Antonelli:2019ytb}
A.~Antonelli, A.~Buonanno, J.~Steinhoff, M.~van de Meent and J.~Vines,
Phys. Rev. D \textbf{99}, no.10, 104004 (2019)
doi:10.1103/PhysRevD.99.104004
[arXiv:1901.07102 [gr-qc]].

\bibitem{Cristofoli:2019neg}
A.~Cristofoli, N.~E.~J.~Bjerrum-Bohr, P.~H.~Damgaard and P.~Vanhove,
Phys. Rev. D \textbf{100}, no.8, 084040 (2019)
doi:10.1103/PhysRevD.100.084040
[arXiv:1906.01579 [hep-th]].

\bibitem{KoemansCollado:2019ggb}
A.~Koemans Collado, P.~Di Vecchia and R.~Russo,
Phys. Rev. D \textbf{100}, no.6, 066028 (2019)
doi:10.1103/PhysRevD.100.066028
[arXiv:1904.02667 [hep-th]].

\bibitem{Maybee:2019jus}
B.~Maybee, D.~O'Connell and J.~Vines,
JHEP \textbf{12}, 156 (2019)
doi:10.1007/JHEP12(2019)156
[arXiv:1906.09260 [hep-th]].




\bibitem{Bjerrum-Bohr:2019nws}
N.~Bjerrum-Bohr, A.~Cristofoli, P.~H.~Damgaard and H.~Gomez,
``Scalar-Graviton Amplitudes,''
JHEP \textbf{11}, 148 (2019)
doi:10.1007/JHEP11(2019)148
[arXiv:1908.09755 [hep-th]].





\bibitem{Naculich:2015zha}
S.~G.~Naculich,
``CHY representations for gauge theory and gravity amplitudes with up to three massive particles,''
JHEP \textbf{05}, 050 (2015)
doi:10.1007/JHEP05(2015)050
[arXiv:1501.03500 [hep-th]].



\bibitem{Cachazo:2013gna}
  F.~Cachazo, S.~He and E.~Y.~Yuan,
  ``Scattering equations and Kawai-Lewellen-Tye orthogonality,''
  Phys.\ Rev.\ D {\bf 90}, no. 6, 065001 (2014)
  doi:10.1103/PhysRevD.90.065001
  [arXiv:1306.6575 [hep-th]].

\bibitem{Cachazo:2013hca}
  F.~Cachazo, S.~He and E.~Y.~Yuan,
  ``Scattering of Massless Particles in Arbitrary Dimensions,''
  Phys.\ Rev.\ Lett.\  {\bf 113}, no. 17, 171601 (2014)
  doi:10.1103/PhysRevLett.113.171601
  [arXiv:1307.2199 [hep-th]].

\bibitem{Cachazo:2013iea}
  F.~Cachazo, S.~He and E.~Y.~Yuan,
  ``Scattering of Massless Particles: Scalars, Gluons and Gravitons,''
  JHEP {\bf 1407}, 033 (2014)
  doi:10.1007/JHEP07(2014)033
  [arXiv:1309.0885 [hep-th]].

\bibitem{Cachazo:2014nsa}
  F.~Cachazo, S.~He and E.~Y.~Yuan,
  ``Einstein-Yang-Mills Scattering Amplitudes From Scattering Equations,''
  JHEP {\bf 1501}, 121 (2015)
  doi:10.1007/JHEP01(2015)121
  [arXiv:1409.8256 [hep-th]].

\bibitem{Cachazo:2014xea}
  F.~Cachazo, S.~He and E.~Y.~Yuan,
  ``Scattering Equations and Matrices: From Einstein To Yang-Mills, DBI and NLSM,''
  JHEP {\bf 1507}, 149 (2015)
  doi:10.1007/JHEP07(2015)149
  [arXiv:1412.3479 [hep-th]].





\bibitem{Gomez:2016bmv}
  H.~Gomez,
  ``$\Lambda$ scattering equations,''
  JHEP {\bf 1606}, 101 (2016)
  doi:10.1007/JHEP06(2016)101
  [arXiv:1604.05373 [hep-th]].

\bibitem{Cardona:2016bpi}
  C.~Cardona and H.~Gomez,
  ``Elliptic scattering equations,''
  JHEP {\bf 1606}, 094 (2016)
  doi:10.1007/JHEP06(2016)094
  [arXiv:1605.01446 [hep-th]].

\bibitem{Bjerrum-Bohr:2018lpz}
  N.~E.~J.~Bjerrum-Bohr, P.~H.~Damgaard and H.~Gomez,
  ``New Factorization Relations for Yang Mills Amplitudes,''
  Phys.\ Rev.\ D {\bf 99}, no. 2, 025014 (2019)
  doi:10.1103/PhysRevD.99.025014
  [arXiv:1810.05023 [hep-th]].

\bibitem{Gomez:2018cqg}
  H.~Gomez,
  ``Scattering equations and a new factorization for amplitudes. Part I. Gauge theories,''
  JHEP {\bf 1905}, 128 (2019)
  doi:10.1007/JHEP05(2019)128
  [arXiv:1810.05407 [hep-th]].

\bibitem{Bjerrum-Bohr:2018jqe}
  N.~E.~J.~Bjerrum-Bohr, H.~Gomez and A.~Helset,
  ``New factorization relations for nonlinear sigma model amplitudes,''
  Phys.\ Rev.\ D {\bf 99}, no. 4, 045009 (2019)
  doi:10.1103/PhysRevD.99.045009
  [arXiv:1811.06024 [hep-th]].

\bibitem{Gomez:2019cik}
  H.~Gomez and A.~Helset,
  ``Scattering equations and a new factorization for amplitudes. Part II. Effective field theories,''
  JHEP {\bf 1905}, 129 (2019)
  doi:10.1007/JHEP05(2019)129
  [arXiv:1902.02633 [hep-th]].






\bibitem{Feng:2019cbe}
  B.~Feng, X.~Li and K.~Zhou,
  ``Expansion of Einstein-Yang-Mills theory by differential operators,''
  Phys.\ Rev.\ D {\bf 100}, no. 12, 125012 (2019)
  doi:10.1103/PhysRevD.100.125012
  [arXiv:1904.05997 [hep-th]].

\bibitem{Hu:2019qdq}
  S.~Q.~Hu and K.~Zhou,
  ``Expansion of tree amplitudes for EM and other theories,''
  arXiv:1907.07857 [hep-th].

\bibitem{Zhou:2019mbe}
  K.~Zhou,
  ``Unified web for expansions of amplitudes,''
  JHEP {\bf 1910}, 195 (2019)
  doi:10.1007/JHEP10(2019)195
  [arXiv:1908.10272 [hep-th]].


\bibitem{Stieberger:2016lng}
  S.~Stieberger and T.~R.~Taylor,
  ``New relations for Einstein-Yang-Mills amplitudes,''
  Nucl.\ Phys.\ B {\bf 913}, 151 (2016)
  [arXiv:1606.09616 [hep-th]].


\bibitem{Schlotterer:2016cxa}
  O.~Schlotterer,
  ``Amplitude relations in heterotic string theory and Einstein-Yang-Mills,''
  JHEP {\bf 1611}, 074 (2016)
  [arXiv:1608.00130 [hep-th]].


\bibitem{Chiodaroli:2017ngp}
  M.~Chiodaroli, M.~Gunaydin, H.~Johansson and R.~Roiban,
  ``Explicit Formulae for Yang-Mills-Einstein Amplitudes from the Double Copy,''
  JHEP {\bf 1707}, 002 (2017)
  doi:10.1007/JHEP07(2017)002
  [arXiv:1703.00421 [hep-th]].

\bibitem{DelDuca:1999rs}
  V.~Del Duca, L.~J.~Dixon and F.~Maltoni,
  ``New color decompositions for gauge amplitudes at tree and loop level,''
  Nucl.\ Phys.\ B {\bf 571}, 51 (2000)
  doi:10.1016/S0550-3213(99)00809-3
  [hep-ph/9910563].


\bibitem{Nandan:2016pya}
  D.~Nandan, J.~Plefka, O.~Schlotterer and C.~Wen,
  ``Einstein-Yang-Mills from pure Yang-Mills amplitudes,''
  JHEP {\bf 1610}, 070 (2016)
  [arXiv:1607.05701 [hep-th]].


\bibitem{delaCruz:2016gnm}
  L.~de la Cruz, A.~Kniss and S.~Weinzierl,
  ``Relations for Einstein-Yang-Mills amplitudes from the CHY representation,''
  Phys.\ Lett.\ B {\bf 767}, 86 (2017)
  [arXiv:1607.06036 [hep-th]].

\bibitem{Fu:2017uzt}
  C.~H.~Fu, Y.~J.~Du, R.~Huang and B.~Feng,
  ``Expansion of Einstein-Yang-Mills Amplitude,''
  JHEP {\bf 1709}, 021 (2017)
  [arXiv:1702.08158 [hep-th]].


\bibitem{Teng:2017tbo}
  F.~Teng and B.~Feng,
  ``Expanding Einstein-Yang-Mills by Yang-Mills in CHY frame,''
  JHEP {\bf 1705}, 075 (2017)
  [arXiv:1703.01269 [hep-th]].

\bibitem{Du:2017kpo}
  Y.~J.~Du and F.~Teng,
  ``BCJ numerators from reduced Pfaffian,''
  JHEP {\bf 1704}, 033 (2017)
  [arXiv:1703.05717 [hep-th]].


\bibitem{Du:2017gnh}
  Y.~J.~Du, B.~Feng and F.~Teng,
  ``Expansion of All Multitrace Tree Level EYM Amplitudes,''
  JHEP {\bf 1712}, 038 (2017)
  [arXiv:1708.04514 [hep-th]].



\bibitem{Cheung:2017ems}
  C.~Cheung, C.~H.~Shen and C.~Wen,
  ``Unifying Relations for Scattering Amplitudes,''
  JHEP {\bf 1802}, 095 (2018)
  doi:10.1007/JHEP02(2018)095
  [arXiv:1705.03025 [hep-th]].


\bibitem{Zhou:2018wvn}
  K.~Zhou and B.~Feng,
  ``Note on differential operators, CHY integrands, and unifying relations for amplitudes,''
  JHEP {\bf 1809}, 160 (2018)
  [arXiv:1808.06835 [hep-th]].

\bibitem{Bollmann:2018edb}
  M.~Bollmann and L.~Ferro,
  ``Transmuting CHY formulae,''
  JHEP {\bf 1901}, 180 (2019)
  [arXiv:1808.07451 [hep-th]].


\bibitem{Bern:2008qj}
  Z.~Bern, J.~J.~M.~Carrasco and H.~Johansson,
  Phys.\ Rev.\ D {\bf 78}, 085011 (2008)
  [arXiv:0805.3993 [hep-ph]].

\bibitem{Chiodaroli:2014xia}
  M.~Chiodaroli, M.~Günaydin, H.~Johansson and R.~Roiban,
  JHEP {\bf 1501}, 081 (2015)
  doi:10.1007/JHEP01(2015)081
  [arXiv:1408.0764 [hep-th]].

\bibitem{Johansson:2015oia}
  H.~Johansson and A.~Ochirov,
  JHEP {\bf 1601}, 170 (2016)
  doi:10.1007/JHEP01(2016)170
  [arXiv:1507.00332 [hep-ph]].

\bibitem{Johansson:2019dnu}
  H.~Johansson and A.~Ochirov,
  JHEP {\bf 1909}, 040 (2019)
  doi:10.1007/JHEP09(2019)040
  [arXiv:1906.12292 [hep-th]].


\bibitem{Lam:2019mfk}
C.~Lam,
``Off-shell Yang-Mills amplitude in the Cachazo-He-Yuan formalism,''
Phys. Rev. D \textbf{100}, no.4, 045009 (2019)
doi:10.1103/PhysRevD.100.045009
[arXiv:1905.05101 [hep-th]].


\bibitem{Zhou:2020umm}
K.~Zhou and G.~J.~Zhou,
``Transmuting off-shell CHY integrals in the double-cover framework,''
[arXiv:2006.12188 [hep-th]].






\end{thebibliography}
\end{document}